\documentclass[a4paper,11pt,amsmath,amssymb]{article} 
\usepackage{jheppub}
\usepackage{amsmath}
\usepackage{array,relsize,float}
\usepackage{pstricks}
\usepackage{color} 
\usepackage{amssymb}
\usepackage{slashed}
\usepackage{tikz-cd}
\usepackage{graphicx} 
\usepackage{epsfig}
\usepackage{multicol} 
\usepackage{hyperref}
\usepackage{pdfpages}
\usepackage{array}
\usepackage{tikz-cd}
\usepackage{amsfonts,layout,appendix,subfigure}
\allowdisplaybreaks
\input{epsf} 

\def\beq{\begin{equation}} \def\eeq{\end{equation}}
\def\beqn{\begin{eqnarray}} \def\eeqn{\end{eqnarray}}
 \def\to{\rightarrow}
\def\nn{\nonumber}

\def\beq{\begin{equation}}
\def\eeq{\end{equation}}
\def\bea{\begin{eqnarray}}
\def\eea{\end{eqnarray}}
\def\beqn{\begin{eqnarray}} \def\eeqn{\end{eqnarray}}
\def\beeq{\begin{eqnarray}}
\def\eeeq{\end{eqnarray}}

\def\nn{\nonumber}

\def\real{{\mathbb{R}}}
\def\complex{{\mathbb{C}}}

\pdfoutput=1

\newcommand{\valencia}{Instituto de F\'{\i}sica Corpuscular, Universitat de Val\`{e}ncia -- Consejo Superior de Investigaciones Cient\'{\i}ficas, Parc Cient\'{\i}fic, E-46980 Paterna, Valencia, Spain.}
\newcommand{\culiacan}{Facultad de Ciencias F\'isico-Matem\'aticas, Universidad Aut\'onoma de Sinaloa, Ciudad Universitaria, CP 80000 Culiac\'an, M\'exico.}
\newcommand{\berlin}{Deutsches Elektronensynchrotron DESY, Platanenallee 6, D–15738 Zeuthen, Germany.}
\newcommand{\munich}{Max-Planck-Institut f\"ur Physik, Werner-Heisenberg-Institut, 80805 M\"unchen, Germany.}

\begin{document}

\title{Mathematical properties of nested residues and their application to
multi-loop scattering amplitudes}
\author[a]{J. Jes\'us Aguilera-Verdugo,}
\author[b]{Roger J. Hern\'andez-Pinto,}
\author[a]{Germ\'an Rodrigo,}
\author[a,c]{German~F. R.~Sborlini} 
\author[a,d]{and William~J.~Torres~Bobadilla}
\affiliation[a]{\valencia}
\affiliation[b]{\culiacan}
\affiliation[c]{\berlin}
\affiliation[d]{\munich}

\emailAdd{jesus.aguilera@ific.uv.es}
\emailAdd{roger@uas.edu.mx}
\emailAdd{german.rodrigo@csic.es}
\emailAdd{german.sborlini@desy.de}
\emailAdd{torres@mpp.mpg.de}

\preprint{IFIC/20-30; DESY 20-172; MPP-2020-184}

\abstract{The computation of multi-loop multi-leg scattering amplitudes plays a key role to improve the precision of theoretical predictions for particle physics at high-energy colliders. In this work, we focus on the mathematical properties of the novel integrand-level representation of Feynman integrals, which is based on the Loop-Tree Duality (LTD). We explore the behaviour of the multi-loop iterated residues and explicitly show, by developing a general formal proof for the first time, that contributions associated to displaced poles are cancelled out. 
The remaining residues, called nested residues as originally introduced in Ref.~\cite{Verdugo:2020kzh}, encode the relevant physical information and are naturally mapped onto physical configurations associated to nondisjoint on-shell states. By going further on the mathematical structure of the nested residues, we prove that unphysical singularities vanish, and show how the final expressions can be written by using only causal denominators. In this way, we provide a mathematical proof for the all-loop formulae presented in Ref. \cite{Aguilera-Verdugo:2020kzc}. }


\setcounter{page}{1}
\maketitle


\section{Introduction}
\label{sec:introduction}
Quantum Field Theories (QFT) have shown to be one of the most successful theoretical constructions to describe the behaviour of Nature at sub-atomic scales. In order to extract reliable predictions from them, it is necessary to develop powerful methods inspired by mathematical ideas. 
The perturbative approach is one of these techniques, and nowadays it stands as the most important framework for high-energy particle physics. In this context, higher-order contributions in the perturbative expansion imply more precise and accurate predictions, with a reduced dependence on the factorization and renormalization scales. Then, it is crucial to compute such corrections in order to explore any small 
discrepancy with the highly-precise experimental data provided by high-energy colliders.

The computation of higher-order contributions in perturbative QFT is not straightforward. The main bottleneck is related to the calculation of virtual contributions, which involves dealing with multi-loop multi-leg Feynman integrals. In the context of current phenomenological studies~\cite{
Abada:2019lih,Abada:2019zxq,Benedikt:2018csr,Abada:2019ono,Blondel:2019vdq,Banerjee:2020tdt}, 
the calculation of scattering amplitudes involving one-loop Feynman integrals~\cite{Ellis:2007qk,vanHameren:2010cp}
is automated in different frameworks~\cite{Berger:2008sj,Cascioli:2011va,Badger:2012pg,Cullen:2014yla,Actis:2016mpe,Alwall:2014hca}. This great achievement is known as the Next-to-Leading order revolution. 
The same automation is not currently attainable for the evaluation
of two- and, hence, multi-loop scattering amplitudes. 
The main obstacles, within standard and traditional approaches, rely on the reduction to the so-called master integrals, by Integration-by-parts identities~\cite{Chetyrkin:1981qh,Laporta:2001dd}, 
and the evaluation of multi-loop Feynman integrals (whose closed formulae are not
all known as in the one-loop case). 

Furthermore, new ideas based on novel mathematical insights are being explored to overcome these limitations. Besides sector decomposition~\cite{Binoth:2000ps,Smirnov:2008py,Carter:2010hi,Borowka:2017idc} and semi-numerical techniques~\cite{Francesco:2019yqt,Bonciani:2019jyb,Czakon:2008zk}, symbolic strategies are being investigated~\footnote{We refer the interested reader to the very complete review~\cite{Heinrich:2020ybq}.}. In particular, the multi-loop integrand reduction algorithm~\cite{Mastrolia:2011pr,Badger:2012dp,Zhang:2012ce,Mastrolia:2012an,Mastrolia:2012wf,Ita:2015tya,Mastrolia:2016dhn,Ossola:2006us}, based on algebraic geometry, decomposes scattering amplitudes in terms of independent integrals. Thus, it is possible to by-pass the tensor reduction and elaborate on a modified implementation of the unitarity based methods~\cite{Peraro:2016wsq,Badger:2017jhb,Abreu:2017hqn,Badger:2019djh}. On top of the studies of decomposition at integrand level, different representations of Feynman integrals~\cite{Baikov:1996rk,Frellesvig:2017aai} promoted the use of algebraic geometry~\cite{Larsen:2015ped,Bern:2017gdk,Zeng:2017ipr,Boehm:2017wjc,Boehm:2018fpv,Bendle:2019csk} and intersection theory~\cite{Mastrolia:2018uzb,Frellesvig:2019kgj,Frellesvig:2019uqt,Weinzierl:2020xyy} to perform a reduction of a multi-loop amplitude to master integrals.


In this paper, we mathematically elaborate on a promising alternative representation of multi-loop amplitudes and Feynman integrals. This approach is based on the Loop-Tree Duality (LTD) theorem~\cite{Catani:2008xa,Bierenbaum:2010cy,Bierenbaum:2012th},
whose main aim is to open loop amplitudes into non-disjoint tree-level amplitudes. 
The genuine LTD theorem is valid in an arbitrary coordinate system. In specific applications, though, it is naturally defined in the Euclidean space of the spacial components of the loop momenta which is implemented by integrating out the energy components.
Several calculations of scattering amplitudes at one-~\cite{Buchta:2015wna,Driencourt-Mangin:2017gop,Jurado:2017xut,Driencourt-Mangin:2019yhu,Plenter:2019jyj,Plenter:2020lop} 
and 
two-loop~\cite{Driencourt-Mangin:2019aix} have been provided within this formalism, as well as the numerical evaluation of multi-loop Feynman integrals up to four loops~\cite{Aguilera-Verdugo:2020kzc,Ramirez-Uribe:2020hes}, which are based on the multi-loop LTD representation recently proposed in Ref.~\cite{Verdugo:2020kzh}. In this context, the LTD approach is currently drawing the attention as a novel tool aimed at overcoming many of the current bottlenecks, and alternative representations have been presented by other authors~\cite{Runkel:2019yrs,Runkel:2019zbm,Capatti:2019ypt,Capatti:2019edf,Capatti:2020ytd}.

Besides the explicit calculation of loop integrals, LTD is the fundamental component of the Four-Dimensional Unsubtraction (FDU) framework~\cite{Hernandez-Pinto:2015ysa,Sborlini:2016gbr,Sborlini:2016hat}, which aims at a simultaneous computation of real and virtual contributions directly in the four physical space-time dimensions. Since loop integrals are re-expressed as phase-space ones, the infrared and threshold singularities are clearly identified in a compact region of the integration domain~\cite{Buchta:2014dfa,Aguilera-Verdugo:2019kbz}. 
Thus, offering the possibility to implement alternative regularization strategies based on the local cancellation of physical singularities present in loop and real-emission contributions~\cite{Becker:2010ng,Becker:2012aqa,Pittau:2012zd,Donati:2013iya,Fazio:2014xea,Soper:1999rd,Soper:1999xk,Soper:2001hu,Kramer:2002cd}.

In this paper, we will focus on the algorithmic methodology to derive compact dual integrand-level representations of scattering amplitudes. Following the ideas presented in a Ref.~\cite{Verdugo:2020kzh}, we  elucidate in more details all the mathematical concepts that are behind the multi-loop LTD representation.
Furthermore, the conjectures originally presented are proven in the present work. 
In order to do so, we follow the application of the Cauchy residue theorem in succession
at multi-loop level together with a short-hand notation that makes clear
simplifications and proofs within this new framework. In this manner, this work represents the necessary mathematical basis to keep developing the program on the application of LTD to three~\cite{Verdugo:2020kzh,Aguilera-Verdugo:2020kzc}, four loops~\cite{Ramirez-Uribe:2020hes} and beyond.

The outline of the paper is the following. In order to properly handle the ideas of the LTD theorem, we present in Sec.~\ref{sec:multiLTD} an overview of the mathematical properties required to define the dual integrand for any scattering amplitude. For this purpose, we explain the concept of the iterated residue on a set of primitive variables, and we give some relations among them. In Sec.~\ref{sec:notation}, we establish a useful notation for the multi-variable iterated residue of a meromorphic function, in order to simplify the symbolic manipulation of the expressions and take advantage of the properties behind this methodology. Once the notation is established, 
we perform the connection with the usual QFT formalism in Sec.~\ref{sec:connectionQFT}. There, we motivate a practical definition of the topological classification of diagrams; in particular, 
we recall the multi-loop configurations  presented in Refs.~\cite{Verdugo:2020kzh,Aguilera-Verdugo:2020kzc}, 
Maximal Loop Topology (MLT), Next-to-Maximal Loop Topology (NMLT) and Next-to-Next-to-Maximal Loop Topologies (N$^2$MLT). Later, we give some examples of the usage of this notation and computational algorithms to derive a formal proof of the all-order formulae presented in Ref.~\cite{Verdugo:2020kzh}. We put special emphasis on highlighting the reduction of complex topologies (i.e. N$^2$MLT and higher) into nested convolutions of MLT ones, which also allows to explain the causal structure of the final compact results. Finally, in Sec.~\ref{sec:conclusions}, we summarize this work and present an outlook of current developments, and potential future applications. Some detailed derivations are provided in the Appendices.


\section{Multi-iterated residues and the Loop-Tree Duality}
\label{sec:multiLTD}
In this section, we establish the mathematical basis of the iterated residue approach for the multi-loop dual representation in the context of the Loop-Tree Duality formalism. 
Then, to begin the discussion, we will consider variables, functions and their pole structure, trying to identify the properties of the multi-variable residue. 
For this reason, we also introduce the physical concepts from the mathematical formalism, in order to appreciate the generality of the presentation. 

We start by studying a multi-variable rational function $f:\real^L \to \complex$. We will restrict the analysis to those rational functions involving quadratic polynomials in the denominator. 
Hence, we consider the following function
\beq
f(\vec{x}) = \frac{{\cal N}(\vec{x})}{(x_1^2-y_1^2)^{\gamma_1}\ldots (x_L^2-y_L^2)^{\gamma_L}\,(z_{L+1}^2-y_{L+1}^2)^{\gamma_{L+1}}\ldots (z_{m}^2-y_{m}^2)^{\gamma_{m}} } \, ,
\label{eq:DefinicionF}
\eeq
where \begin{equation}
    z_l=k_l+\sum\limits_{j=1}^{L}\beta_j^{(l)}x_j, \,\qquad l\in\{L+1,\hdots,m\},
\end{equation} with $\beta\in\{-1,0,1\}$, $k_l\in\real$, $\vec{x}=(x_1,\ldots,x_L)\in\real^L$ is the \emph{primitive} set of variables and $y_i \in \complex$, for every $i\in\{1,\ldots,m\}$, are the roots of the different polynomials that factorize the denominator. This functional form is sufficient to capture the essential mathematical structure of multi-loop multi-leg scattering amplitudes involving quadratic propagators and standard Feynman rules for the interaction vertices. As explained in Sec.~\ref{sec:connectionQFT}, external momenta are encoded as shifts in the pole structure along the real axis. 

Keeping in mind the physical motivation, we restrict to the case $\gamma_i \in \mathbb{N}$ which corresponds to having elements that strictly belongs to the denominator of the whole rational function. By convention, we take $m \geq L+1$, and
\beq
z_{L+1}=-\sum_{j=1}^L x_j + k_{L+1} \, ,
\label{eq:DefinicionZ}
\eeq
with $k_{L+1} \in \real$, an arbitrary number. Regarding the structure of the numerator, ${\cal N}$, we will consider polynomial functions in $\real[\vec{x}]$ without any other restriction. 
It is worth appreciating that most of the proofs presented throughout 
this paper require very weak constraints on ${\cal N}$, because this guarantees the generality of the approach and allows to extend their validity to almost any QFT.

Once the function $f$ is described, we now move to the computation of its integral over the whole domain, $\real^L$. Explicitly,
\beq
I = \left( \prod_{i=1}^L \, \int \frac{dx_i}{2\pi \imath} \right) \, f(\vec{x}) \, ,
\label{eq:DefinicionI}
\eeq
where the so-called primitive variables play the role of integration variables. Due to the rational structure indicated in Eq.~(\ref{eq:DefinicionF}), the most natural strategy to perform the integration consists in the application of Cauchy's Residue Theorem (CRT). In order to use this theorem, we must assume that $f$ fulfils integrability on $\real^L$; in other words, the integral in Eq.~(\ref{eq:DefinicionI}) must exist. Once integrability is guaranteed, we can:
\begin{itemize}
\item use Fubini's theorem to change the integration order freely; 
\item change integration variables (through linear combinations, re-scaling, etc.);
\item iterate the application of CRT.
\end{itemize} 
At this point, the motivation for inquiring into this computation is related to the definition of multi-loop scattering amplitudes. As mentioned before, we can consider $f$ as the integrand originated in any one-dimensional perturbative QFT and the integral in Eq.~(\ref{eq:DefinicionI}) as the associated amplitude. Moreover, a clever redefinition of variables will be enough to extend the treatment to QFT in an arbitrary number of space-time dimensions.

In order to make use of CRT, let us take a look at the singularities of $f$. From the meromorphic structure of the function $f$ (inspired on the integrands obtained in Feynman representation), it is seen that it only contains poles which arise as the solution of the equation,
\beq\label{eqn:4}
\left( \prod_{i=1}^{L} (x_i^2-y_i^2)^{\gamma_i} \right )\, \times \, \left( \prod_{j=L+1}^{m} (z_j^2-y_j^2)^{\gamma_j} \right) = 0 \, ,
\eeq
for the primitive variables. Notice that some solutions might have multiplicity higher than one, leading to multiple poles. The solutions of Eq.~(\ref{eqn:4}) for $x_1$ are given by,
\beq\label{eqn:5}
x_{1,j} \in \mathrm{Poles}[f,x_1]=\left\{ \pm y_1, \pm y_{L+1}-k_{L+1} - x_2 -\ldots - x_{L}, \ldots, \pm y_{l}-k_l - \sum\limits_{j=1}^{L} \beta^{(l)}_j x_j  \right\} ,
\eeq for some $l\in\{L+2,\hdots,m\}$.
Furthermore, when applying CRT, the integration variables are extended to the complex plane. 
Then, we can extend all of them simultaneously, or one-by-one. Also, we must take special care with those parameters entering in Eq.~(\ref{eq:DefinicionI}) that might displace the pole position in the complex plane. 
Thus, with the purpose of removing these ambiguities, we consider that:
\begin{itemize}
\item the primitive variables are extended to the complex plane successively, but not simultaneously, which implies that when computing the residue for $x_i$, we promote $x_i \in \complex$ but we keep $x_j \in \real$ for any $i \neq j$;
\item every $y_i$ has positive real part and an infinitesimally small negative imaginary part.
\end{itemize}
Regarding the last point, we assume,
\beq\label{eqn:6}
y_i \to \tilde{y}_i = \sqrt{y_i^2 - \imath 0} \, ,
\eeq
with the purpose of defining the \emph{complex prescription}. Again, this is inspired by physical concepts, and coincides with the customary $+\imath 0$ Feynman prescription introduced in QFT calculations. For the sake of simplicity, and to avoid overloading the notation, here and in the following, we drop the tilde, i.e. $y_i \equiv \tilde{y}_i$.

More in details, the algorithmic procedure of the \textit{iterated residue} begins with the promotion of the primitive variable $x_i$ to $\complex$, through the natural inclusion mapping $i:\complex^{\real^L}\hookrightarrow\complex^{\complex\times\real^{L-1}}$, where $i(f(\vec{x}))=f(\vec{x})$. Then, the residue of the latter function is computed along a contour included in the half-plane with $\mathrm{Im}(x_i)<0$ with the function $\mathrm{Res}:\complex^{\complex\times\real^{L-1}}\rightarrow\complex^{\real^{L-1}}$. It is mandatory to say that the residue is well defined because the arguments of the function contain only one complex variable, $x_i$. In this way, the iterated residue algorithm can be understood as an iterated application of the functor,
 \begin{equation}\label{eqn:10}
    \mathrm{Res}\circ i:\complex^{\real^{L}}\to\complex^{\real^{L-1}},
\end{equation} 
and, thus, the iterated residue is represented by
\begin{equation}\label{eq:Diagrama1}
\begin{tikzcd}
\complex^{\real^{L}}\arrow{r}{\mathrm{Res}\,\circ\,i}&\complex^{\real^{L-1}}\arrow{r}{\mathrm{Res}\,\circ\,i}&\cdots\arrow{r}{\mathrm{Res}\,\circ\,i}&\complex.
\end{tikzcd}
\end{equation}

In the following, we will explore the consequences of these prescriptions. In particular, we will prove that some contributions vanish because of non-trivial cancellations related with the quadratic dependence of the denominators. As a consequence, we will explicitly show that the integral in Eq.~(\ref{eq:DefinicionI}) can be computed by just looking at the residues of specific poles.

\subsection{Cancellation of residues from displaced poles}
\label{ssec:noncausal}
Now, we can choose one by one the integration variables and apply CRT on Eq.~(\ref{eq:DefinicionI}). 
We  follow the natural order, i.e., $x_1$ first, $x_2$ then, and so on. 
The final result for the integral $I$ in Eq.~(\ref{eq:DefinicionI}) is independent of the ordering, but intermediate expressions and integrand level results  exhibit a non-trivial dependence on it. 
Then, to illustrate this behavior,
we start by applying CRT in $x_1$: we promote $x_1 \in \real \to \complex$ and close the integration contour from the lower part of the complex plane, i.e.
\beq
I = -\left(\prod_{i=2}^{L} \int \frac{d x_i}{2\pi \imath}\right) \,  \sum_ {x_{1,j} \in \mathrm{Poles}[f,x_1]} {\rm Res}\left(f(\vec{x}),\{x_1,x_{1,j}\}\right) \, \theta(-{\rm Im}(x_{1,j})) \, ,
\label{eq:ResiduoPaso1}
\eeq
where we assume that the function $f$ fulfils the good-convergence hypothesis that allows integrability at infinity. The Heaviside theta function selects only those poles with negative imaginary part. Due to the fact that $x_i \in \real \ \forall i \neq 1$ and ${\rm Im}(y_l)<0$, it turns out that the only poles from (\ref{eqn:5}) that contribute are
\beq
\mathrm{Poles}^{(+)}[f,x_1]=\left\{y_1, y_{L+1}-k_{L+1}-x_2 -\ldots - x_{n-1}, \ldots, y_l -k_l- \sum_j \beta^{(l)}_j x_j \right\} , 
\label{eqn:8}
\eeq for $l\in\{L+1,\hdots,m\}$.
Then, Eq.~(\ref{eq:ResiduoPaso1}) reduces to
\beq
I = -\left(\prod_{i=2}^{L} \int \frac{d x_i}{2\pi \imath}\right) \,  \sum_ {x_{1,j} \in \mathrm{Poles}^{(+)}[f,x_1]} {\rm Res}\left(f(\vec{x}),\{x_1,x_{1,j}\}\right) \, ,
\label{eq:ResiduoPaso1simple}
\eeq
without any loss of generality. 

The next step consists in performing the second iterated integral in the primitive variable $x_2$. This is the crucial step, because here it is necessary to recalculate the position of the poles in $x_2$. After considering the poles in the variable $x_1$, the position of some poles in the variable $x_2$ are displaced. At this point, and without giving the explicit formulae of the resulting integrand in each application of CRT, our purpose is to show that only the residues of specific poles contribute to the final result. 
We shall now select all the poles in the variable $x_2$ with negative imaginary part. Since $\{y_i\}_{i=1,\ldots,m}$ are the only parameters with non-vanishing imaginary part, the selection of the poles is done according to the specific combinations of $y_i$'s that might appear as arguments of Heaviside theta functions. 

In order to clarify the meaning of the last sentence, we present an explicit example. 
We assume that $f$ takes the form
\beq
f(\vec{x}) = \frac{1}{(x_1^2-y_1^2) \ldots (x_L^2-y_L^2) \, (z_{L+1}^2-y_{L+1}^2)} \, ,
\label{eq:Fsimple}
\eeq
and we compute the poles in $x_1$. By selecting only those with negative imaginary part, we have
\beq
\mathrm{Poles}^{(+)}[f,x_1] = \{y_1, y_{L+1}-k_{L+1}-x_2-\ldots -x_L \} \, ,
\label{eq:Omega1EJEMPLO}
\eeq
and the sum of residues is given by
\beqn
\nn {\rm Res}(f,\{x_1,{\rm Im}(x_{1})<0\}) &=& \sum_{x_{1,j} \in \mathrm{Poles}^{(+)}[f,x_1]} {\rm Res}(f,\{x_1,x_{1,j}\})
\\ \nn &=& \frac{1}{2y_1\,(x_2^2-y_2^2) \ldots (x_L^2-y_L^2) \, ((y_1+x_2+\ldots+x_L-k_{L+1})^2-y_{L+1}^2)}
 \\ \nn &+& \frac{1}{2y_{L+1}\,((y_{L+1}+k_{L+1}-x_2-\ldots-x_L)^2-y_1^2)(x_2^2-y_2^2) \ldots (x_L^2-y_L^2)} \, .
 \\
 \label{eq:FirstResidue}
\eeqn
Then, by applying CRT on $x_2$, the possible poles will be
\beq\label{eqn:14}
\mathrm{Poles}[f,x_1;x_2] = \{\pm y_2, \pm y_1 + y_{L+1} - x_3 - \ldots - x_L+k_{L+1},\pm y_{L+1} - y_1  - x_3 - \ldots - x_L +k_{L+1}\} \, ,
\eeq
and, from this set, we must only retain those with negative imaginary part. Performing the complete computation, 
we obtain
\begin{align}
{\rm Res}&(\,{\rm Res}(f,\{x_{1},{\rm Im}(x_{1})<0\})  \,,\{x_{2},{\rm Im}(x_{2})<0\})\nonumber \\
 & =\sum_{x_{2,l}\in\mathrm{Poles}[f,x_1,x_2]}{\rm Res}(\,({\rm Res}(f,\{x_{1},{\rm Im}(x_{1})<0\}),\{x_{2},x_{2,l}\}\,)\,\theta(-{\rm Im}(x_{2,l}))\,,
\label{eq:EjemploThetasCancel0}
\end{align}
where we can identify two kinds of contributions:
\begin{itemize}
\item two terms associated to $x_2=y_2$ and $x_2=y_1+y_{L+1}-x_3-\ldots-x_L+k_{L+1}$, which read
\begin{align}
\nn  {\rm Res}(\,({\rm Res}&(f,\{x_1,{\rm Im}(x_{1})<0\}),\{x_2,y_{2}\}\,) 
\\ \nn =& \frac{1}{4y_1 y_2\,(x_3^2-y_3^2)\ldots (x_L^2-y_L^2)((y_1+y_2+x_3+\ldots+x_L-k_{L+1})^2-y_{L+1}^2)} 
\\ +& \frac{1}{4y_{L+1} y_2\,((y_{L+1}-y_2-x_3-\ldots-x_L+k_{L+1})^2-y_1^2)\ldots (x_L^2-y_L^2)} \, ,
\label{eq:EjemploThetasCancel1a}
\end{align}
and
\begin{align}
\nn  {\rm Res}(\,{\rm Res}&(f,\{x_1,{\rm Im}(x_{1})<0\}),\{x_2,y_1+y_{L+1}-x_3-\ldots-x_L+k_{L+1}\}\,) 
\\  =& \frac{1}{4y_1 y_3\,((y_1+y_{L+1}-x_3-\ldots-x_L+k_{L+1})^2-y_2^2)(x_3^2-y_3^2)\ldots (x_L^2-y_L^2)} \, ,
\label{eq:EjemploThetasCancel1b}
\end{align}
\item and one contribution which contains a non-trivial theta function, i.e.
\begin{align}
\nn  &\left[ {\rm Res}(\,{\rm Res}(f,\{x_1,y_1\}),\{x_2,y_{L+1}-y_1-x_3-\ldots-x_L+k_{L+1}\}\,) \right.
\\ \nn +&  {\rm Res}(\,{\rm Res}(f,\{x_1,y_{L+1}-x_2-\ldots-x_L+k_{L+1}\}),\\
&\left.\phantom{\mathrm{Res}(}\{x_2,y_{L+1}-y_1-x_3-\ldots-x_L+k_{L+1}\}\,) \right] \, \theta({\rm Im} (y_1-y_{L+1})) \, .
\label{eq:EjemploThetasCancel2}
\end{align}
\end{itemize}
It is crucial to appreciate that, after the explicit computation, the sum of the iterated residues in Eq.~(\ref{eq:EjemploThetasCancel2}) vanishes. 
These are the contributions associated to what we call \emph{displaced poles}, whose position in the complex plane depend on the evaluation of residues in the previous variable. 
Moreover, in this example, the indices $1$ and $2$ for the integrated primitive variables were arbitrary; this implies that the result can be extended by induction to any number of primitive variables. Besides that, we would like to highlight that this property holds for any function $f$ as described in Eq.~(\ref{eq:DefinicionF}), because it is a consequence of the quadratic pole structure. The master formula responsible of the cancellation is
\begin{align}
\nn  {\rm Res}({\rm Res}(F(x_i,&x_j),\{x_i,y_i+a_i\}),\{x_j,y_k-y_i+a_{ij}-a_i\}) 
\\ =-& {\rm Res}\left({\rm Res}\left(F(x_i,x_j),\{x_i,y_k-x_j+a_{ij}\}\right),\{x_j,y_k-y_i+a_{ij}-a_i\}\right) \,  ,
\label{eq:MasterCancellation}
\end{align}
with
\beq
F(x_i,x_j) =  \frac{P(x_i,x_j)}{((x_i-a_i)^2-y_i^2)^{\gamma_i}((x_i+x_j-a_{ij})^2-y_k^2)^{\gamma_k}} \, ,
\label{eq:FuncionEspecial}
\eeq
where $a_i$ and $a_{ij}$ are linear combinations of $k_l$'s or $y_l$'s, excluding the primitive variables $x_i$ and $x_j$, and $P$ is any meromorphic function without poles for the variables $\{x_i,x_j\}$ in the location indicated in Eq.~(\ref{eq:MasterCancellation}). Notice that Eq.~(\ref{eq:MasterCancellation}) is valid for poles of arbitrary order. The proof goes through a direct computation of the iterated residue from the Laurent series of the function $F(x_i,x_j)$. A formal proof of this expression is available in Appendix~\ref{app:Cancelacion}. This is the first mathematical proof of the cancellation of the displaced poles contributions and is the main result of this section.

After each iteration the resulting function will have the form given in Eq.~(\ref{eq:FuncionEspecial}), so we conclude that Eq.~(\ref{eq:MasterCancellation}) implies the cancellation of all the non-trivial Heaviside theta functions. Since this result holds step-by-step in the calculation, the final expression will also be free of residues associated to displaced poles. To conclude this example, we notice that when evaluating Eqs.~(\ref{eq:EjemploThetasCancel1a}) and~(\ref{eq:EjemploThetasCancel1b}) for the case $L=2$, in Eq.~(\ref{eq:DefinicionZ}), we obtain extremely compact expressions, namely,
\begin{align}
\nn & {\rm Res}(\,{\rm Res}(f,\{x_1,{\rm Im}(x_{1})<0\}),\{x_2,{\rm Im}(x_{2})<0\}\,)
\\ \nn &= \frac{1}{4y_1 y_2\,((y_1+y_2-k_{3})^2-y_{3}^2)} 
 + \frac{1}{4y_2 y_3\, ((y_3+y_1+k_3)^2-y_2^2)}\\
 &+ \frac{1}{4y_1 y_3\, ((y_3-y_2+k_3)^2-y_1^2)} \nn
\\ &= -\frac{1}{8y_1 y_2 y_3}\ \left( \frac{1}{y_1+y_2+y_3-k_3}+\frac{1}{y_1+y_2+y_3+k_3}\right) \, .
\label{eq:ResultadoEjemplito}
\end{align}
Moreover, we can appreciate that the denominator of the last expression in Eq.~(\ref{eq:ResultadoEjemplito}) only involve sums of $y_i$'s. This observation is very important, since it restricts the presence of certain kind of singularities at integrand level that can be interpreted in terms of causality~\cite{Aguilera-Verdugo:2020kzc}. We will provide explicit all-loop order proofs that support the achievement of this causal structure in a very general family of topologies.

As the expressions obtained after the computation of the iterated residue are free of contributions from displaced poles, it is fair to ignore them. Thus, the remaining contributions are associated to what we call \textit{nested residues}~\cite{Aguilera-Verdugo:2020kzc}.

\subsection{Towards a geometrical and physical interpretation}
\label{ssec:interpretation}
In the previous example, we observed two interesting properties, namely:
\begin{enumerate}
\item Cancellation of the residues from displaced poles: \emph{residues of poles whose imaginary part depends on different sign combinations of $y_i$'s  cancel.}
\item Cancellation of non-causal contributions: \emph{only denominators involving sums of $y_i$'s survive after adding up all the terms produced by the nested residues.}
\end{enumerate}
Regarding the first item, it means that terms proportional to
\beq
\theta\left({\rm Im}\left(y_j - y_i \right)\right)
\eeq
cancel as successive evaluation of the corresponding residues leads to terms with opposite signs. These displaced poles are located in the upper or in the lower part of the complex plane depending on the specific value of the $y_i$'s. On the contrary, the remaining contributions are those involving same-sign-combinations of $y_i$'s. These poles always remain on one side of the real axis. 

\begin{figure}[H]
    \centering
    \includegraphics{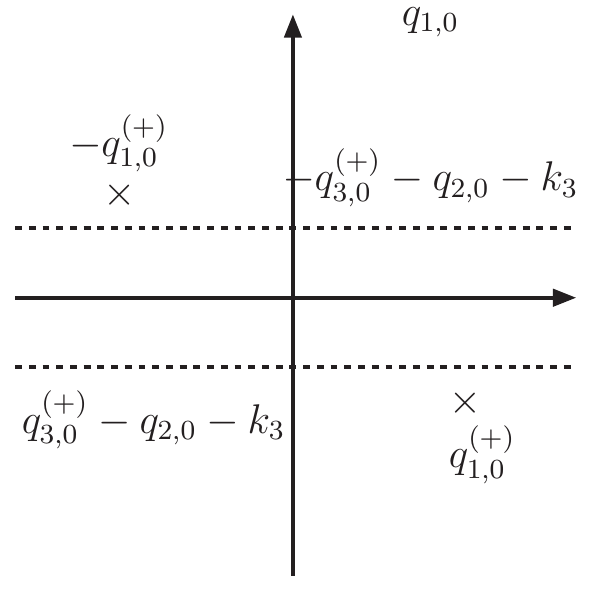}
    \caption{Pole structure of a rational function of two variables.}
    \label{fig:poles1}
\end{figure}

A geometrical interpretation of this cancellation is as follows. Let us start with the function \begin{equation}\label{eqn:example}
    f(x_1,x_2)=\frac{1}{(x_1^2-y_1^2)(x_2^2-y_2^2)((x_1+x_2+k_3)^2-y_3^2)}.
\end{equation} 
The poles of the function in Eq.~(\ref{eqn:example}), in the variable $x_1$, are located at \begin{equation}
    \mathrm{Poles}[f,x_1]=\{\pm y_1,\pm y_3-x_2-k_3\},
\end{equation} as it is shown in Fig.~\ref{fig:poles1}. It is important to notice that $x_1$ has been extended to $\complex$, while $x_2$ is still considered as a real parameter. Thus the poles $\pm y_3-x_2$ are located along horizontal lines, depending on the value of $x_2$.

\begin{figure}[H]
    \centering
    \includegraphics[width=0.8\textwidth]{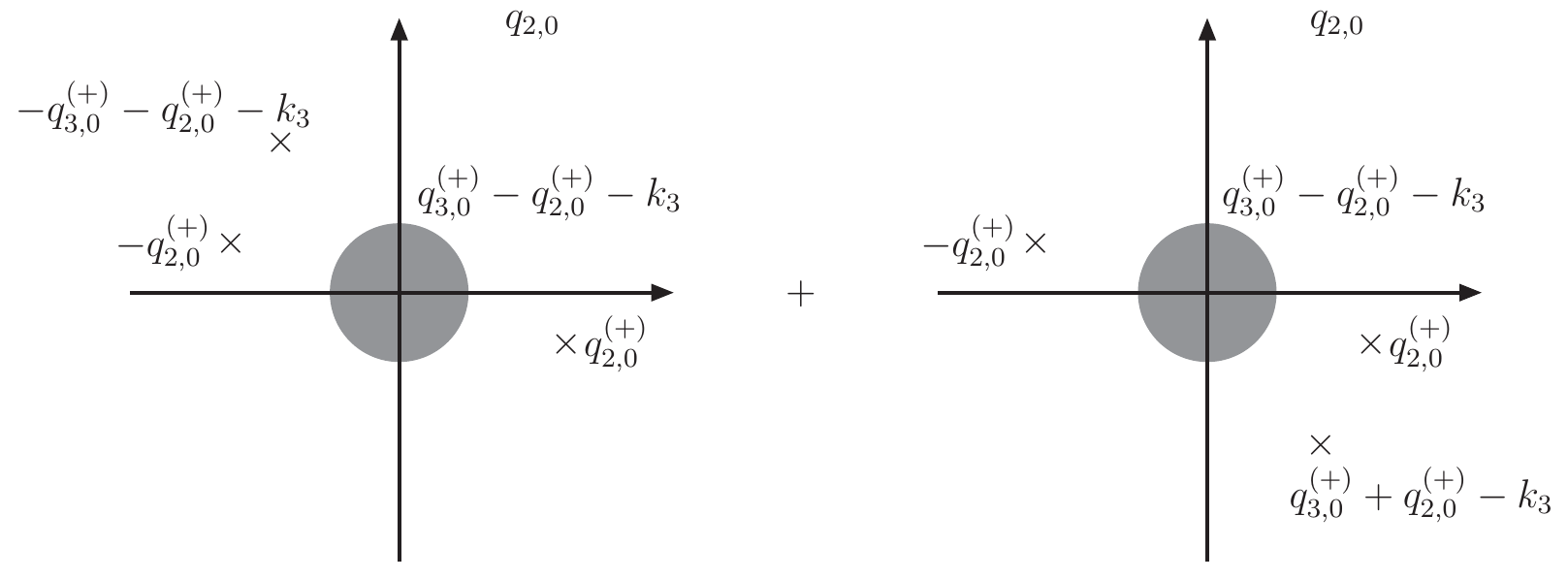}
    \caption{Pole structure of the first residue of a rational function of two variables.}
    \label{fig:poles2}
\end{figure}

Computing the residue of this function, closing the contour on the lower half-plane, the enclosed poles are \begin{equation}
    \mathrm{Poles}^{(+)}[f,x_1]=\{y_1,y_3-x_2-k_3\},
\end{equation} each of which gives the corresponding residue, \begin{equation}\begin{split}
    \mathrm{Res}[f(x_1,x_2),\{x_1,y_1\}]&=\frac{1}{2y_1(x_2^2-y_2^2)((y_1+x_2+k_3)^2-y_3^2)},\\
    \mathrm{Res}[f(x_1,x_2),\{x_1,y_3-x_2-k_3\}]&=\frac{1}{2y_3((y_3-x_2-k_3)^2-y_1^2)(x_2^2-y_2^2)}.
\end{split}\label{eqn:resexamp1}\end{equation}
Thus, it is obtained \begin{equation}
    \begin{split}
        \mathrm{Res}[f(x_1,x_2),\mathrm{Im}(x_1)<0]&=\mathrm{Res}[f(x_1,x_2),\{x_1,y_1\}]+\mathrm{Res}[f(x_1,x_2),\{x_1,y_3-x_2-k_3\}]\\
        &=\frac{1}{2y_1(x_2^2-y_2^2)((y_1+x_2+k_3)^2-y_3^2)}\\
        &+\frac{1}{2y_3((y_3-x_2-k_3)^2-y_1^2)(x_2^2-y_2^2)}.
    \end{split}\label{eqn:resextot}
\end{equation} The poles of the first term are located at \begin{equation}
    \mathrm{Poles}[f,x_1,x_2]=\{\pm y_2,\pm y_3-y_1-k_3\},
\end{equation} and the poles of the second term are at \begin{equation}
    \mathrm{Poles}[f,x_1,x_2]=\{\pm y_2,y_3\pm y_1-k_3\}.
\end{equation} Diagrammatically, the poles structure of each term in Eq.~(\ref{eqn:resextot}) are depicted in Fig.~\ref{fig:poles2}, where particularly the pole $y_{3}-y_{2}$ is located somewhere inside the grey circle. This is because the imaginary part can be positive or negative, depending on the explicit values of $y_{3}$ and $y_{2}$.

Since in this example we are dealing with simple poles, the computation of the residue is straightforward, although the overall interpretation that follows is also valid for multiple poles. In the first line of Eq.~(\ref{eqn:resexamp1}), the function must be evaluated in $x_1\to y_1$ and, in the second line, the evaluation is performed in $x_1\to y_3-x_2-k_3$. On top of that, we notice that the factor in the denominator is symmetric under the transformation $x_2\to-x_2$. As it is shown in Fig.~\ref{fig:poles2}, the location of the poles, except in the case $y_{3}-y_{2}-k_3$, is symmetric with respect to the origin. This can be interpreted as a connection between the first and the second term through the transformation $x_2\to-x_2$. This situation admits a graphical interpretation: performing the integration through the contour selected in the second term (on the lower half-plane) is equivalent to choose the integration contour on the upper half-plane for the first term. The subtlety here is that the pole $y_{3}-y_{2}-k_3$ will lead to a vanishing contribution because it appears inside both contours and thus, produces two contributions with opposite signs. To be more explicit, the imaginary part of the displaced poles could be positive or negative. On one hand, if the imaginary part of the displaced pole is positive, then it does not belong to the interior of any of the integration contours. On the other hand, if the imaginary part of the displaced pole is negative, through the transformation $x\to-x$ (which can be interpreted as a rotation in $\pi$) a new pole will appear in $x_2\to-y_3+y_2+k_3$, leading to a relative minus sign between the contributions of the displaced poles.

From the physical point of view, the cancellation of displaced poles is a consequence of the causal structure of multi-loop Feynman integrals, since these contributions do not have a representation in terms of cut diagrams. As discussed in previous studies~\cite{Catani:2008xa,Verdugo:2020kzh,Aguilera-Verdugo:2019kbz}, evaluating the residue of the integrand expression for a given loop diagram is equivalent to set on shell certain internal lines. The LTD representation is in fact independent of the loop momentum labelling and the ordering in the computation of the iterated residues. However, individual terms, associated to displaced poles, exhibit an explicit dependence on the loop labelling and ordering through the argument of non-trivial Heaviside theta functions. In consequence, they do not contribute to the final result, as we proved in the Sec.~\ref{ssec:noncausal}.

Up to now, we justified the cancellation of the residues of displaced poles relating them to unphysical contributions. In the same spirit, we can think about same-sign-combinations of $y_i$'s as aligned contributions. The sign of the imaginary part of the poles is directly related with the energy flow of the internal propagators that are being set on shell. Thus, this means that only those contributions associated to a properly-aligned energy flow will remain. In the physics language, these are \emph{causal} configurations and are directly related to the threshold singularities of scattering amplitudes in the context of the LTD approach, as discussed in Ref.~\cite{Aguilera-Verdugo:2019kbz}.

\section{Symbolic treatment of iterated residues}
\label{sec:notation}
Once the mathematical basis of the nested multi-residue strategy was explained, we aim at simplifying the symbolic treatment of expressions. For this purpose, we  deepen into the development of a physically-inspired notation that captures the main features of the iterated residue method and allows for a straightforward implementation.

We start from the conventions introduced in Sec.~\ref{sec:multiLTD}. When studying the analytic properties of the function given in Eq.~(\ref{eq:DefinicionF}), we can drop the information related to the specific variables and keep only the associated indices. In this way, we introduce the following notation: \begin{equation}\label{eqn:21}
    \begin{split}
        F_i(j)&:=\left\{\begin{array}{cl}
            \left(x_j^2-y_i^2\right)^{-\gamma_i} & \forall j\in\{1,\hdots ,L\}\ \forall i\in\{1,\hdots ,m\} \\
            \left(z_j^2-y_i^2\right)^{-\gamma_i} & \forall j\in\{L+1,\hdots ,m\}\ \forall i\in\{1,\hdots ,m\}
        \end{array}\right.\\
        F(i_1,i_2,\hdots ,i_k)&:=\mathcal{N}(\vec{x})\prod\limits_{j=1}^{k}F_{i_j}(i_j)\ \ \forall k\in\{1,\hdots ,m\}.
    \end{split}
\end{equation}
It is important to notice that there are no common zeroes between $F_i(i)^{-1}$ and $F_j(j)^{-1}$ when $i\neq j$. Thus, by using the definitions of Eq.~(\ref{eqn:21}), the original function in Eq.~(\ref{eq:DefinicionF}) can be rewritten as \begin{equation}\label{eqn:22}
    f(\vec{x})=F(1,2,\hdots .,m).
\end{equation}
From now on we consider the case ${\cal N}=1$, but we remark that most of the results remain unchanged as the concept of the residue does not depend on the specific structure of the given function. Instead, it depends only on the pole structure (i.e. in the zeroes of denominators).
 
This notation has some interesting properties which allows to simplify the presentation of explicit results. In particular: \begin{enumerate}
    \item Having two or more indices together in the $k$-th argument is interpreted as summing the variables associated to those indices inside the propagator corresponding to $y_k$. For instance,\begin{equation}\label{eqn:24}
        F(1,2,12)=\frac{1}{(x_1^2-y_1^2)^{\gamma_1}(x_2^2-y_2^2)^{\gamma_2}((x_1+x_2+k_3)^2-y_3^2)^{\gamma_3}}
    \end{equation}
    where $12 \equiv x_1+x_2$.
    \item Having a bar above a given index means that the associated variable is inverted, $x_i\to-x_i$. E.g.
    \begin{equation}\label{eqn:25}
        F(1,2,1\overline{2})=\frac{1}{(x_1^2-y_1^2)^{\gamma_1}(x_2^2-y_2^2)^{\gamma_2}((x_1-x_2+k_3)^2-y_3^2)^{\gamma_3}}
    \end{equation}
    \item Having sub-indices within a given argument means that associated $y$-parameters are added or subtracted according to the bar convention. For example,\begin{equation}\label{eqn:26}
        F(1_2,2_{\overline{1}},\overline{2})=\frac{1}{((x_1+y_2)^2-y_1^2)^{\gamma_1}((x_2-y_1)^2-y_2^2)^{\gamma_2}((-x_2+k_3)^2-y_3^2)^{\gamma_3}}.
    \end{equation}
    \item Having 0 as one of the arguments represents that the corresponding primitive variables where replaced by the combination of $y$-parameters present in the sub-indices. Also, if 0 is in the $k$-th argument, and if it has as sub-index $(k)$, it means that it has been computed the residue in the poles within the associated factor. E.g.,\begin{equation}\label{eqn:27}
        F(0_{(1)},2)=\mathrm{Res}\left(\left(x_1^2-y_1^2\right)^{-\gamma_1}\left(x_2^2-y_2^2\right)^{-\gamma_2},\{x_1,y_1\}\right),
    \end{equation}
    which corresponds to computing the residue in $x_1=y_1$. It is important to emphasize that, for a function such as $F(1,2,12)$, the sub-index $3$, representing a term $y_3$, is different from the sub-indices $12$, representing the sum $y_1+y_2$.
\end{enumerate}

Some immediate properties of the quadratic structure of the functions $F_i$ are 
that $F_i(j_k)=F_i(\overline{j}_{\overline{k}})$ and $F_i(\overline{j}_k)=F_i(j_{\overline{k}})$, and thus, it is possible to fix the overall sign of a given variable. Also, it is straightforward that $\overline{\overline{k}}=k$ and $k\overline{k}=0$. This two properties allows us to compute efficiently the nested residues.

\subsection{Efficient residue computation}
\label{ssec:simpleresidue}
At this point, we are interested in computing efficiently the residues of a generic function with the form $F(1,2,\hdots,m)$. Moreover, we aim to use algebraic and symbolic properties to avoid computing unnecessary terms which cancel in each iteration of the residue. For illustrative reasons, we  focus the discussion on the case of functions with the particular form,
\begin{equation}\label{eqn:30}
    F(1,2,\hdots,L+1) \ .
\end{equation}
In the following, we will use the equivalent notations $L+1\equiv\overline{1\hdots L} = -1-2-\hdots-L$ with the purpose of shorten the presentation of results, wherever the expressions are unambiguously defined. The poles with negative imaginary part of $F$, in the complex variable $x_1$, are
\beq
\mathrm{Poles}^{(+)}[F,x_1]=\left\{ y_1, y_{L+1}-\sum\limits_{i=2}^{L}x_i+k_{L+1} \right\} \, .
\eeq
Whence, for the computation of the residue at $x_1=y_1$, we obtain, 
\begin{equation}\label{eqn:31}
    \begin{split}
        \mathrm{Res}(F(1,\hdots,L+1),\{x_1,y_1\})&=F(0_{(1)},2,\hdots ,L,2\hdots L_1),
    \end{split}
\end{equation}
whilst the residue in the other pole is given by \begin{equation}\label{eqn:32}
    \begin{split}
        \mathrm{Res}\left(F(1,\hdots ,L+1),\left\{x_1,y_{L+1}-\sum\limits_{i=2}^{L}x_i+k_{L+1}\right\}\right)&=F(\overline{2\hdots L}_{L+1},2,\hdots ,L,0_{(L+1)})\\
        &=F(2\hdots L_{\overline{L+1}},2,\hdots ,L,0_{(L+1)}).
    \end{split}
\end{equation}
In general, after computing a residue, we obtain different functions of the form $F_i(i)$, $F_i(j)$, $F_i(j_k)$ and $F_i(j_{\overline{k}})$. The set of negative imaginary part poles associated with these functions are 
\begin{equation}\label{eqn:33}
    \begin{split}
        \mathrm{Poles}^{(+)}[F(i)]&=\mathrm{Poles}[F_i(j)]=\{y_i+k_{L+1}\},\\
        \mathrm{Poles}[F_i(j_k)]&=\{y_i-y_k+k_{L+1}\},\\
        \mathrm{Poles}[F_i(j_{\overline{k}})]&=\{y_k+y_i+k_{L+1},y_k-y_i+k_{L+1}\}.
    \end{split}
\end{equation}
As already noticed in Sec.~\ref{sec:multiLTD}, the iterated residue might lead to expressions whose poles are not always within the integration contour (i.e. with negative imaginary part); the so-called \emph{displaced poles}. For instance, $y_k-y_i+k_{L+1}$ might have a positive or negative imaginary part, depending on the specific values of the $y$-parameters. Thus, it is necessary to impose the condition $\mathrm{Im}(y_k-y_i)<0$ when computing the residue with the function \begin{equation}
    \theta(i\overline{k}):=\theta(\mathrm{Im}(y_k-y_i)).
\end{equation} 
Hence, 
\begin{equation}\label{eqn:34}
    \begin{split}
        \mathrm{Res}(F(i),\{x_i,y_i\})&=\mathrm{Res}((x_i^2-y_i^2)^{-\gamma_i}, \{x_i,y_i\})\equiv F_i(0_{(i)}),\\
        \mathrm{Res}(F_i(j),\{x_j,y_i\})&=\mathrm{Res}((x_j^2-y_i^2)^{-\gamma_i}, \{x_j,y_i\})\equiv F_i(0_{(i)})\\
        \mathrm{Res}(F_i(j_k),\{x_j,y_i-y_k\})&=\mathrm{Res}(((x_j+y_k)^2-y_i^2)^{-\gamma_i}, \{x_j,y_i-y_k\})\equiv\theta(\overline{i}k)F_i(0_{(i)}),\\
        \mathrm{Res}(F_i(j_{\overline{k}}),\{x_j,y_i+y_k\})&=\mathrm{Res}(((x_i-y_k)^2-y_i^2)^{-\gamma_i}, \{x_j,y_i+y_k\})\equiv F_i(0_{(i)}),\\
        \mathrm{Res}(F_i(j_{\overline{k}}),\{x_j,y_k-y_i\})&=\mathrm{Res}(((x_i-y_k)^2-y_i^2)^{-\gamma_i}, \{x_i,y_k-y_i\})\equiv-\theta(\overline{k}i)F_i(0_{(i)}).
    \end{split}
\end{equation}

Remarkably, from Eq.~(\ref{eqn:34}), we directly appreciate the cancellation of the contributions associated to displaced poles, due to the appearance of a relative minus sign. In consequence, this cancellation becomes explicit for the second and subsequent iteration of residues. The formal proof presented in Appendix~\ref{app:Cancelacion} is based on identifying the contributions with opposite signs that cancel among them. Notice that similar cancellations have been observed in~\cite{Capatti:2019ypt,Capatti:2019edf} by considering poles with positive and negative imaginary parts for specific configurations without a formal general proof as presented in this paper.

In order to clarify the notation, we present an explicit example. For instance, 
\begin{equation}
    \label{eqn:35}\begin{split}
        F(1,2,12)&\rightarrow F(0_{(1)},2,2_1)+F(2_{\overline{3}},2,0_{(3)})\\
        &\rightarrow F(0_{(1)},0_{(2)},0_{12})+F(0_{(1)},0_{13},0_{(3)})+F(0_{2\overline{3}},0_{(2)},0_{(3)})\,,
    \end{split}
\end{equation} where the arrow represents the computation of the residue of function on the left. In the first line, the residue in the variable $x_1$ originates two terms. Then, in the second line, we identified all the poles in $x_2$ associated to the expression in line 1, and we computed the residues. By putting the sub-indices and subtracting from the main index, we identify the variable in which we apply CRT and the pole where we evaluate. The extra contribution $F(0_{(1)},0_{1\overline{3}},0_{(3)})$ has not been considered because it corresponds to a displaced pole.

A practical way to see this procedure is as follows. The computation of the first residue in Eq.~(\ref{eqn:35}) corresponds to the poles $x_1=y_1$ and $x_1+x_2=y_3+k_3$, or, in this notation, corresponds to $1=0_1$ and $12=0_3$. These two index equations are equivalent to $\overline{1}_1=0$ and $\overline{12}_3=0$, respectively, and, as these expressions are 0, they can be added or subtracted in other arguments in the corresponding iteration of the residue. 
Thus, for the computation of the first residue, in the case of the pole $x_1=y_1$, it can be obtained $12=12(\overline{1}_1)=(1\overline{1})2_1=2_1$. Analogously, for the pole $x_1=y_3-x_2+k_3$, $1=1(\overline{12}_3)=(1\overline{1})\overline{2}_3=\overline{2}_3$. 
Finally, as $\overline{2}_3$ becomes the first argument of the function, it represents a factor of the form $((-x_2+y_3+k_3)^2-y_1^2)^{-\gamma_1}=((x_2-y_3-k_3)^2-y_1^2)^{-\gamma_1}$, it is possible to change its sign, this is, $\overline{2}_3\to\overline{(\overline{2}_3)}=\overline{\overline{2}}_{\overline{3}}=2_{\overline{3}}$.

It is important to highlight that, through the computation of the iterated residue, the displaced poles can clearly be identified in two cases. In the first case, the displaced poles can be seen as those arriving from arguments with at least one sub-index without bar. 
For instance, in Eq.~(\ref{eqn:35}), after the computation of the first iteration of the iterated residue, with respect to the variable $x_1$, in the first term, $F(0_{(1)},2,2_1)$, the poles associated to the third argument $2_1$ are one positive-imaginary-part pole (which is outside the integration contour) and one displaced pole. 
The second case corresponds to the arguments with all sub-indices with bar. In this case, there is one negative-imaginary-part pole and one displaced pole. Then, the displaced pole is identified as the one leaving the argument with at least one sub-index without bar and at leas one sub-index with bar. 
For instance, in Eq.~(\ref{eqn:35}), after the computation of the first iteration of the iterated residue, with respect to $x_1$, the second term $F(2_{\overline{3}},2,0_{(3)})$ has its first argument $2_{\overline{3}}$, and thus contributes with one displaced pole (located in $x_2=y_3-y_1+k_3$, or, in this notation, $2=0_{\overline{1}3}$) and a negative-imaginary-part pole (located in $x_2=y_1+y_3+k_3$, or, in this notation $2=0_{13}$). 
In this manner, the nested residue can be computed directly by considering just the poles associated with the arguments without sub-indices and the poles associated with the arguments with all its sub-indices with bar, whenever the index of the integration variable does not have a bar.

\subsection{Recursive representation with nested residues}
\label{ssec:recursion}
In Refs.~\cite{Verdugo:2020kzh,Aguilera-Verdugo:2020kzc}, we showed very compact formulae for some Feynman diagram topologies at all-loop orders. We provide in this paper formal proofs of the validity of these expressions, by taking advantage of the notation previously introduced with the purpose of unveiling the recursive relations that naturally manifest when computing the nested residues. This will lead to inductive proofs of the beforehand mentioned formulae.

In the following, we explore some relations among nested residues to identify the potential recursive structures. We sequentially calculate the residue in the primitive variables, and simplify the result of each step to find the dependence on the number of iterations. Thus, we infer the functional forms that we obtain after the $i$-th iteration and proof the inductive step.

So, let us study some examples to find the recursions. The simplest case corresponds to the application of the iterated residue to a function $F(1,\hdots,L)$, with independent arguments. This is the case of
\beq
F(1,2)=\frac{1}{(x_1^2-y_1^2)^{\gamma_1}(x_2^2-y_2^2)^{\gamma_2}}\,,
\eeq
which corresponds to the class of \emph{factorizable} functions and does not deserve further comments. Here, we discuss about the non-trivial case of \emph{non-factorizable} functions, and the most symmetric example is found when there are only two dependent arguments in each step of the iterated residue, as for the function $F(1,\hdots,L+1)$. We start by noticing an interesting property when computing the first $i$-th nested residue for this function, \begin{equation}
    \begin{split}
        &F(1,\hdots,L+1)\rightarrow F(1,\ldots,L-i)\\
        &\times\sum\limits_{j=L-i+1}^{L+1}F(0_{(L-i+1)},\hdots,0_{(j-1)},1\hdots(L-i)_{(L-i+1)\hdots (j-1)\overline{(j+1)\hdots(L+1)}},0_{(j+1)},\hdots,0_{(L+1)}),
    \end{split}
\end{equation} where $F(1,\hdots,L-i)$ was factorized because it does not depend on the primitive variables that are involved in the computation of the iterated residues in the last $i$-th variables. Whence, it is straightforward that, after the computation of all the iterated residues, we obtain \begin{equation}\label{eqn:37}
    \begin{split}
        F(1&,\hdots,L+1)\rightarrow\sum\limits_{i=1}^{L+1}F(0_{(1)},\hdots,0_{(i-1)},0_{1\hdots (i-1)\overline{(i+1)\hdots(L+1)}},0_{(i+1)},\hdots,0_{(L+1)}).
    \end{split}
\end{equation}In Appendix~\ref{app:IterativeFormula}, we provide a formal proof of these relations for simple poles, by using a more physically-inspired notation and, in Appendix~\ref{app:alebra}, we prove the generalization for multiple poles. Again, we recall that the arrow indicates that the expression in the r.h.s. corresponds to the nested residue of the expression in the l.h.s.


Another important relation can be found for functions of the form $F(1,\hdots,L+2)$, where we defined $L+2\equiv\overline{12}=-1-2$. In this case, after a direct computation of all the iterated residues and reordering the result, we obtain
\begin{equation}\label{eqn:38}
    \begin{split}
        F(1,&\hdots,L+2)\\
        &\rightarrow\left[F\left(0_{(1)},0_{(2)},0_{12}\right)+F\left(0_{(1)},0_{1\overline{(L+2)}},0_{(L+2)}\right)+F\left(0_{\overline{2(L+2)}},0_{(2)},0_{(L+2)}\right)\right]\\
        &\times\sum\limits_{j=3}^{L+1}F\left(0_{(3)},\hdots,0_{(j-1)},0_{1\hdots(j-1)\overline{(j+1)\hdots(L+1)}},0_{(j+1)},\hdots,0_{(L+1)}\right)\\
        &+\left[F\left(0_{(1)},0_{1\overline{3\hdots(L+1)}}\right)+F\left(0_{\overline{2\hdots(L+1)}},0_{(2)}\right)\right]F_{L+2}\left(0_{\overline{3\hdots(L+1)}}\right)F\left(0_{(3)},\hdots,0_{(L+1)}\right).
    \end{split}
\end{equation}
The nested residues lead to two terms, with a strong resemblance to the factorization formulae presented in Ref.~\cite{Aguilera-Verdugo:2019kbz}. The different contributions are characterized according to the poles considered for the residue computation. In the following, we will explain better this separation, although we defer to Sec.~\ref{sec:results} the physical interpretation in terms of on-shell internal propagators.

Let us extend the results for more general functions. For a given function $F(1,\hdots,m)$, if $\{1,\hdots,\rho\}$ are the indices of the primitive variables appearing in three or more arguments then, whenever $\gamma_i=1$ and the $k$-parameters vanish for $i> \rho$, direct computation of the first $L-\rho$ residues of the function $F$ leads to \begin{equation}\label{eqn:simp1}
    F(1,\hdots,\rho,\hdots,L+1,\hdots,m)\rightarrow F(1,\hdots,\rho,L+2,\hdots,m)F_{(\rho+1)\hdots(L+1)}(1\hdots\rho^*),
\end{equation} where \begin{equation}\label{eqn:simp2}
    F_{(\rho+1)\hdots(L+1)}(1\hdots\rho^*)=\left(\left(\sum\limits_{k=1}^{\rho}x_k\right)^2-\left(\sum\limits_{k=\rho+1}^{L+1}y_k\right)^2\right)^{-1}.
\end{equation}
The right hand side of Eq.~(\ref{eqn:simp1}) encodes the singular behaviour of the function in the left hand side and the function defined in Eq.~(\ref{eqn:simp2}) represents an auxiliary propagator which summarize the information associated to the sets $\rho+1$ through $L+1$.

From a physical perspective, this expression plays the role of a modified propagator with an alternative on-shell condition. Also, it can be thought as a consequence of applying momentum conservation; we will return to this point later, in Sec.~\ref{sec:results}. Notice that it is enough to show the validity of these expressions for simple poles (the formalization of this claim is given in Appendix~\ref{app:alebra}). The generalization to the non-vanishing $k$-parameters case presents no extra difficulty and is delayed to Sec.~\ref{sec:results}.

Thus, the r.h.s. of Eq.~(\ref{eqn:38}) can be expressed as \begin{equation}\label{eqn:snmlt}
    F(1,\hdots,L+2)\rightarrow F(1,2,L+2)F_{3\hdots(L+1)}(L+2^*),
\end{equation}where \begin{equation}\label{deffalpha}
    F_{3\hdots(L+1)}(12^*)=\left(\left(x_1+x_2\right)^2-\left(\sum\limits_{k=3}^{L+1}y_k\right)^2\right)^{-1}.
\end{equation}
After a direct computation of the iterated residue, we end up with \begin{equation}\label{eqn:snmltfinal}
    \begin{split}
        F(1,\hdots,L+2)&\rightarrow F(1,2,L+2)\otimes F_{3\hdots(L+1)}(L+2^*)\\
        &+F(1,2)\otimes F_{3\hdots(L+1)}(0_{(3\hdots(L+1))}^*)\otimes F(L+2),
    \end{split}
\end{equation}where it is understood that \begin{equation}
    \begin{split}
        F(1,2,L+2)\ \otimes\ &F_{3\hdots(L+1)}(L+2^*)=[F(0_{(1)},0_{1\overline{(L+2)}},0_{(L+2)})\\
        &+F(0_{\overline{2(L+2)}},0_{(2)},0_{(L+2)})]F_{3\hdots(L+1)}(0_{L+2}^*)\\
        &+F(0_{(1)},0_{(2)},0_{12})F_{3\hdots(L+1)}(0_{12}^{*}),\\
        F(1,2)\ \otimes\ &F_{3\hdots(L+1)}(0_{(3\hdots(L+1))}^*)\ \otimes\ F(L+2)\\
        &=[F(0_{(1)},0_{1\overline{3\hdots(L+1)}})+F(0_{\overline{2\hdots(L+1)}},0_{(2)})]\\
        &\times F_{L+2}(0_{3\hdots(L+1)})F_{3\hdots(L+1)}(0_{(3\hdots(L+1))}^*).
    \end{split}\label{eqn:convdefnmlt}
\end{equation}
The convolution symbol, $\otimes$, means that the residues of the different factors involved in the operation are connected. For instance, in the first relation of Eq.~(\ref{eqn:convdefnmlt}), both factors depend on $L+2$. Thus, when computing the residue in the associated pole, both factors will be modified. The result consists in a sum of the different evaluations of residues in the shared poles.

In this discussion, it is important to point out that $F_{3\hdots(L+1)}(0_{L+2}^*)\neq F_{3\hdots(L+1)}(0_{12}^*)$ since, by definition, \begin{equation}
    \begin{split}
        F_{3\hdots(L+1)}(0_{L+2}^{*})&=\frac{1}{y_{L+2}^2-(y_3+\hdots+y_{L+1})^2},\\
        F_{3\hdots(L+1)}(0_{12}^{*})&=\frac{1}{(y_1+y_2)^2-(y_3+\hdots+y_{L+1})^2}.
    \end{split}
\end{equation} In the first case, it is important to notice that it is not possible to consider simultaneously the pole associated to the function $F_{L+2}$, so that the function $F_{L+2}(L+2)=F(L+2)$ is factorized.

We would like to highlight that the great simplification from Eq.~(\ref{eqn:38}) to Eq.~(\ref{eqn:snmltfinal}) is due to the fact that all the information of the function $F(3,\hdots,L+1)$ is encoded within $F_{3\hdots(L+1)}(12^*)$. This is, when the iterated residue of the function in Eq.~(\ref{eqn:snmlt}) is computed for the pole associated with $F_{3\hdots(L+1)}(L+2^*)$, the result is equivalent to take the terms of the iterated residue of the function $F(1,\hdots,L+2)$ where the poles associated with the indices $\{3,\hdots,L+1\}$ are all included. On the contrary, if the pole associated to the function $F_{3\hdots(L+1)}(L+2^*)$ is not considered, then Eq.~(\ref{eqn:simp1}) assures that the function $F_{3\hdots(L+1)}$ equals the sum encoded in the summation symbol of Eq.~(\ref{eqn:38}). In other words, we establish the identity \begin{equation}\begin{split}
    F&_{(\rho+1)\hdots(L+1)}(1\hdots\rho^*)\\
    &\leftrightarrow \sum\limits_{i=\rho+1}^{L+1}F(0_{(\rho+1)},\hdots,0_{(i-1)},1\hdots\rho_{(\rho+1)\hdots(i-1)\overline{(i+1)\hdots(L+1)}},0_{(i+1)},\hdots,0_{(L+1)}).
\end{split}\end{equation} This is, when we compute the residue of an expression including the function $F_{(\rho+1)\hdots(L+1)}$ on its negative imaginary part pole (when it appears $F_{(\rho+1)\hdots(L+1)}(0_{((\rho+1)\hdots(L+1))}^*)$), it can be connected with a more general expression by the substitution \begin{equation}
    F_{(\rho+1)\hdots(L+1)}(0_{((\rho+1)\hdots(L+1))}^*)\leftrightarrow F(0_{(\rho+1)},\hdots,0_{(L+1)}).
\end{equation} Furthermore, after the computation of the nested residue, factors of the form $F_{(\rho+1)\hdots(L+1)}(0_{\beta})$, where $\beta$ is a given combination of the indices $\{1,\hdots,\rho\}$ with an arbitrary bar configuration, can be replaced according to \begin{equation}
    F_{(\rho+1)\hdots(L+1)}(0_{\beta})\leftrightarrow\sum\limits_{i=\rho+1}^{L+1}F(0_{(\rho+1)},\hdots,0_{(i-1)},0_{\beta(\rho+1)\hdots(i-1)\overline{(i+1)\hdots(L+1)}},0_{(i+1)},\hdots,0_{(L+1)}).
\end{equation}

With this identification between the $(L-\rho)$-th iterated residue of the function $F(\rho+1,\hdots,L+1)$ with multiple-poles or non-vanishing $a$-parameters, and the function $F_{(\rho+1)\hdots(L+1)}$ which is its simplification for simple poles and vanishing $a$-parameters, it is possible to interpret the arguments $\{\rho+1,\hdots,L+1\}$ as a single function with only one factor in the denominator of the form $(x^2-y^2)$ with multiplicity 1.

All this discussion is needed to consider the case of a more complex function of the form $F(1,\hdots,L+3)$, where we define $L+3\equiv\overline{23}=-2-3$. In this case, we factorize the sets 4 through $L+1$ and Eq.~(\ref{eqn:simp1}) reduces to \begin{equation}
    \begin{split}
        F(1,\hdots,L+3)\rightarrow F(1,2,3,L+2,L+3)F_{4\hdots(L+1)}(123^*).
    \end{split}
\end{equation}
Because of the complexity of this function, a simple result of the iterated residues is not expected. Still, the final expression can be rewritten in terms of convolutions of simpler functions. We obtain \begin{equation}\label{eqn:3.30}\begin{split}
    F(1,\hdots,L+3)\rightarrow&F(1,2,3,L+2,L+3)\otimes F_{4\hdots(L+1)}(0_{(4\hdots(L+1))}^*)\\
    +&F(1\cup L+3,2,L+2\cup3)\otimes F_{4\hdots(L+1)}(123^*).
\end{split}\end{equation}

In the previous formula, we used the following explicit definition for the convolutions:\begin{equation}
    \begin{split}
        F(1,2,3,&L+2,L+3)\otimes F_{4\hdots(L+1)}(0_{(4\hdots(L+1))}^*)\\
        &=[F(0_{4\hdots(L+1)(L+3)},0_{4\hdots(L+1)\overline{(L+2)}(L+3)},0_{4\hdots(L+1)\overline{(L+2)}},0_{(L+2)},0_{(L+3)})\\
        &+F(0_{(1)},0_{(2)},0_{12\overline{4\hdots(L+1)}},0_{12},0_{1\overline{4\hdots(L+1)}})\\
        &+F(0_{(1)},0_{1\overline{3\hdots(L+1)}},0_{(3)},0_{\overline{3\hdots(L+1)}},0_{1\overline{4\hdots(L+1)}})\\
        &+F(0_{4\hdots(L+1)(L+3)},0_{(2)},0_{2(L+3)},0_{24\hdots(L+1)(L+3)},0_{(L+3)})\\
        &+F(0_{\overline{2\hdots(L+1)}},0_{(2)},0_{(3)},0_{\overline{3\hdots(L+1)}},0_{\overline{24\hdots(L+1)}})\\
        &+F(0_{(1)},0_{1\overline{(L+2)}},0_{\overline{4\hdots(L+1)}},0_{(L+2)},0_{1\overline{4\hdots(L+1)}})\\
        &+F(0_{4\hdots(L+1)(L+3)},0_{\overline{3}(L+3)},0_{(3)},0_{\overline{3\hdots(L+1)}},0_{(L+3)})\\
        &+F(0_{\overline{2(L+2)}},0_{(2)},0_{4\hdots(L+1)\overline{(L+2)}},0_{(L+2)},0_{\overline{2}4\hdots(L+1)\overline{(L+2)}})]F_{4\hdots(L+1)}(0_{(4\hdots(L+1))}^*),
    \end{split}
\end{equation}and\begin{equation}
    \begin{split}
        F(1\cup L+3,&2,L+2\cup3)\otimes F_{4\hdots(L+1)}(123^*)=[F(0_{(1)},0_{1\overline{(L+2)}},0_{(3)},0_{(L+2)},0_{1\overline{3(L+2)}})\\
        &+F(0_{\overline{2(L+2)}},0_{(2)},0_{(3)},0_{(L+2)},0_{2\overline{3}})\\
        &+F(0_{\overline{3(L+2)(L+3)}},0_{\overline{3(L+3)}},0_{(3)},0_{(L+2)},0_{(L+3)})]F_{4\hdots(L+1)}(0_{3(L+2)}^{*})\\
        &+[F(0_{(1)},0_{(2)},0_{2\overline{(L+3)}},0_{12},0_{(L+3)})+F(0_{(1)},0_{3(L+3)},0_{(3)},0_{13(L+3)},0_{(L+3)})\\
        &+F(0_{(1)},0_{1\overline{(L+2)}},0_{1\overline{(L+2)}(L+3)},0_{(L+2)},0_{(L+3)})]F_{4\hdots(L+1)}(0_{1(L+3)}^*)\\
        &+F(0_{\overline{2(L+2)}},0_{(2)},0_{2(L+3)},0_{(L+2)},0_{(L+3)})F_{4\hdots(L+1)}(0_{2(L+2)(L+3)}^*)\\
        &+F(0_{(1)},0_{(2)},0_{(3)},0_{12},0_{23})F_{4\hdots(L+1)}(0_{123}^*).
    \end{split}
\end{equation}

The general case of the expression for the iterated residue of $F(1,\hdots,L+3)$ becomes even more complicated, but the relations of the $F_{4\hdots(L+1)}$ remain. This is,\begin{equation}
    F_{4\hdots(L+1)}(0_{(4\hdots(L+1))}^*)=F(0_{(4)},\hdots,0_{(L+1)}),
\end{equation} and for the rest of the arguments, it is the sum of functions of the form $F(4,\hdots,L+1)$ where the iterated residues have been computed and one of the poles associated to the arguments $\{4,\hdots,L+1\}$ has not been taken into account.

The results obtained in Eqs.~(\ref{eqn:snmltfinal}) and~(\ref{eqn:3.30}) are remarkable because they give a relation between a family of functions with certain complexity and convolutions of simpler ones. This association can be understood in a diagrammatic way, by using the concepts and ideas taken from QFT. We will discuss that in the following section, and apply these relations to provide recursive proofs of some physical results in Sec.~\ref{sec:results}.

We would like to highlight that the discussion of this section was independent of the numerator of the integrand $f$ and independent of the order in which the residues of the poles of $F$ are evaluated, which ensures that it is general enough to allow a straightforward application to scattering amplitudes computations.

\section{Nested residues for scattering amplitudes}
\label{sec:connectionQFT}
In the context of perturbative QFT, we are interested in computing scattering amplitudes, and, in particular, their higher-order representations. These representations are given in terms of loop Feynman diagrams, where internal virtual particles circulate as quantum fluctuations. A diagram with $L$ loops posses $L$ independent or \emph{primitive} loop momenta, $\{ \ell_i \}_{i=1,\ldots,L}$, that define the integration space. The momenta flowing through the internal lines can be grouped into sets, in such a way that all the momenta inside a set $s$ are of the form $q_{i_s} = \ell_s + k_{i_s}$, with $\ell_s$ and $k_{i_s}$ linear combinations of primitive loop and external momenta, respectively. Because of momentum conservation, the number of momentum sets, $n$, is always larger than the number of loops, $n \geq L+1$ for $L \geq 2$ non-factorizable Feynman diagrams.\footnote{In the particular case of one-loop diagrams, all the internal lines depend on a single momenta; thus, there is only one set.}

Likewise, the loop diagrams involve Feynman propagators, whose dependence on the loop momenta settles the pole structure of the whole amplitude. Thus, we introduce the scalar Feynman propagator,
\beq
G_F(q_i) \equiv  \frac{1}{q_i^2-m_i^2+\imath 0} = \frac{1}{q_{i,0}^2-(q_{i,0}^{(+)})^2} \, ,
\eeq
where $q_i$ represents the momentum flow through this line, $m_i$ is the mass of the particle and
\beq
q_{i,0}^{(+)}=\sqrt{\boldsymbol{q}_i^2+m_i^2-\imath0} \, ,
\eeq
is the corresponding positive on-shell energy. The $+\imath 0$ prescription is crucial for establishing the location of the poles in the complex plane, and thus defining the physical modes for the on-shell states. 

In order to write down explicit representations for scattering amplitudes, we need to use the corresponding Feynman rules. In general, an $L$-loop amplitude with $N$ external particles is given by
\beq
{\cal A}_N^{(L)}(1,\ldots,n) = \int_{\ell_1, \ldots, \ell_L} \, {\cal N}(\{\ell_i\}_{L},\{p_j\}_{N}) \times G_F(1,\ldots,n) \, ,
\label{eq:GenericAmplitude}
\eeq
where ${\cal N}$ is a function given by the Feynman rules of the theory that depends on the loop and external momenta, and
\beq
G_F(1,\ldots,n) = \prod_{i \in 1 \cup \ldots\cup n} (G_F(q_i))^{\gamma_i} \, ,
\label{eq:ProductoGF}
\eeq
is a product of Feynman propagators over the union of the $n$ momenta sets, allowing arbitrary positive powers $\gamma_i$ for each line. As usual, the integration measure is defined as
\beq
\int_{\ell_s} = -\imath \mu^{4-d} \int \frac{d^d\ell_s}{(2\pi)^d} \, ,
\eeq
for an arbitrary number $d$ of space-time dimensions. If the loop momenta is decomposed as $\ell_s=(\ell_{s,0},\boldsymbol{\ell}_s)$, Eq.~(\ref{eq:GenericAmplitude}) can be expressed, in a very general form, as
\begin{align}
 \nn  {\cal A}_N^{(L)}&(1,\ldots,n) \\
 &=\int_{\boldsymbol{\ell}_1 \ldots \boldsymbol{\ell}_L} 
\int_{\ell_{1,0} \ldots \ell_{L,0}} \ \frac{{\cal N}(\{\ell_{i,0}\}_{L})}{(\ell_{1,0}^2-(q_{1,0}^{(+)})^2)^{\gamma_1}\ldots ((\sum_j \beta_j \ell_{j,0}+k_{m,0})^2-(q_{m,0}^{(+)})^2)^{\gamma_m}} \, ,
\label{eq:GenericAmplitude2}
\end{align}
where $\{k_{m,0}\}$ are linear combinations of the energies of external momenta and $\mathcal{N}$ is a polynomial in the loop energies.

At this point, the connection with the notation introduced in Secs.~\ref{sec:multiLTD} and~\ref{sec:notation} is straightforward: Eq.~(\ref{eq:GenericAmplitude2}) agrees with the functional form showed in Eqs.~(\ref{eq:DefinicionF}) and~(\ref{eq:DefinicionI}). If the primitive variables $\{x_l\}$ are identified with the energy component of the loop momenta, then the $y_i$ parameters are mapped onto the positive on-shell energies, $q_{i,0}^{(+)}$; and, the real constants $k_j$ are associated with linear combinations of the energy of the external particles $k_{m,0}$. In general, the primitive variables can be identified with any other component of the loop momenta.


Regarding the short-hand notation introduced in Sec.~\ref{sec:notation}, we identify a Feynman propagator associated to a line $i \in s$, $G_F(q_{i_s})$, with $F_{i_s}(i_s)$. The nested residues correspond to the so-called dual amplitudes, as defined in Eqs. (5)-(6) of Ref.~\cite{Verdugo:2020kzh}. Explicitly, we establish the connection
\begin{align}\label{eqn:4.6}
   \nonumber  F(0_{(1)},\hdots,0_{(i-1)},0_{1\hdots(i-1)\overline{(i+1)\hdots L(L+1)}},&0_{(i+1)},\hdots,0_{(L+1) })
    \\ & \rightarrow {\cal A}_D(1,\hdots,i-1,\overline{i+1},\hdots,\overline{L+1};i)\,.
\end{align} We anticipate that this result justifies the so-called MLT formulae presented in Ref.~\cite{Verdugo:2020kzh}, and shall be explained in more detail in Sec.~\ref{sec:results}.

Finally, we would like to make a comment on Eq.~(\ref{eq:GenericAmplitude2}). After the application of the iterated CRT, the original loop amplitude will involve only integrals in the spatial components of the loop momenta, i.e. $\boldsymbol{\ell}_i$, which are inside the definition of $q_{i,0}^{(+)}$. The integration space is now Euclidean, instead of the original Minkowskian one. This fact, together with the compact form of the dual representation, points towards a more efficient numerical implementation within this formalism, as we already tested in Ref.~\cite{Aguilera-Verdugo:2020kzc}.

\subsection{Topological families}
\label{ssec:families}
Scattering amplitudes can be classified according to their internal momentum flow, which translates into specific topological structures for the associated Feynman diagrams. In Ref.~\cite{Verdugo:2020kzh}, we introduced a systematic classification scheme of multi-loop topologies, which includes specific families of diagrams with arbitrary number of loops.

Given an $L$-loop diagram ($L \geq 2$) with $\ n \geq L+1$ sets of internal propagators, we define the topological complexity as $k=n-L$. In this way, the Maximal Loop Topology (MLT), which is the most symmetric configuration, has topological complexity $k=1$ and the Next-to-Maximal Loop Topology (NMLT) has topological complexity $k=2$. In general, a N$^{k-1}$MLT diagram at $L$ loops has topological complexity $k$, and we will denote it N$^{k-1}$MLT$(L)$. 

With all these definitions in mind, we proceed to present explicit results for MLT$(L)$, NMLT$(L)$ and N$^2$MLT$(L)$ configurations in the following sections, focusing on their recursive structure and the decomposition into convolutions of lower-complexity topologies.

\section{Selected results for topological families}
\label{sec:results}
\begin{figure}[htb]
    \centering
    \includegraphics[width=0.3\textwidth]{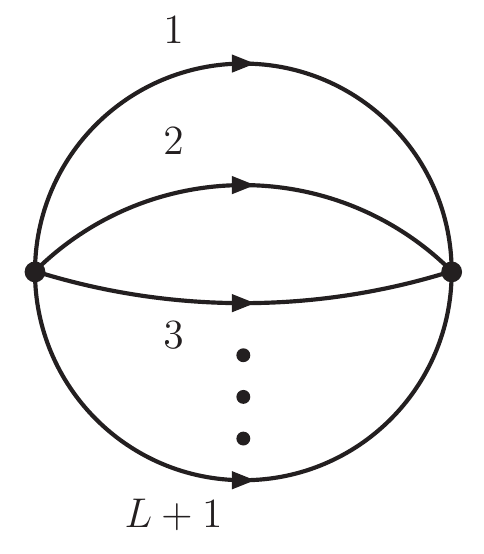}
    \caption{The general form of the MLT$(L)$ topology.}
    \label{fig:MLTdiagramA}
\end{figure}

Multi-loop scattering amplitudes with an arbitrary number of external legs are objects that involve integrands with a structure that can be properly described in terms of the functions $F_i(j)$ defined in Sec.~\ref{sec:notation}. In this section, we make use of their properties shown above, in order to highlight their recursive structure.
The most symmetric loop configuration in any QFT can be encoded into the MLT$(L)$ diagram which is given by
\begin{align}
\mathcal{A}_{\mathrm{MLT}}^{(L)}(1,\hdots,L+1)&\equiv\int\limits_{\ell_1,\hdots,\ell_L}\mathcal{N}(\{\ell_i\}_L,\{p_j\}_N)\times G_F(1,\hdots,L+1)\,,
\label{eqn:mltamp}\end{align}
and is graphically depicted in Fig.~\ref{fig:MLTdiagramA}.

For the moment, we have not discussed anything regarding the structure of the numerator
in Eq.~\eqref{eq:GenericAmplitude2}, however, as discussed in Sec.~\ref{sec:multiLTD},
the treatment at integrand level is independent of the explicit structure of the latter. It is straightforward to notice that the symbolic handling of the expressions relies on the iterated application of CRT, that only requires to indicate the pole location, without making use of the explicit functional form of the numerators. Hence, for the sake of simplicity, we restrict the following discussion to the case $\mathcal{N}=1$,
since all our dual representations can be straightforwardly generalised to any numerator. 
We can also restrict the demonstrations to vacuum diagrams, i.e. those without external particles, because they contain sufficient information regarding the loop-momenta dependence of each internal set of propagators. The generalization to loop configurations with an arbitrary number of externa particles is achieved by implicitly considering the sum over nested residues within each set of propagators. 

\begin{figure}[H]
    \centering
    \includegraphics[width=0.65\textwidth]{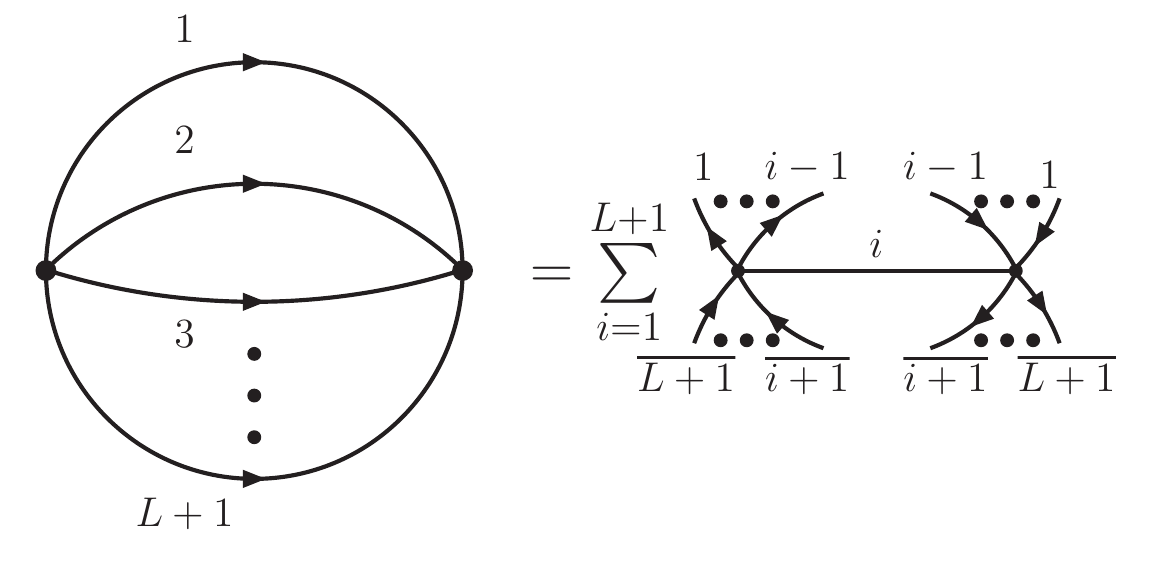}
    \caption{Opening of MLT$(L)$ topology into non-disjoint amplitudes.}
    \label{fig:MLTdual}
\end{figure}

In order to find the LTD realization of Eq.~(\ref{eqn:mltamp}), we just need to interprete 
Eqs.~(\ref{eqn:21}) and~(\ref{eqn:37}) in terms of Feynman and dual propagators as,
\begin{equation}\label{eqn:5.1}
    G_F(1,\hdots,L+1)\rightarrow\sum\limits_{i=1}^{L+1}G_D(0_{(1)},\hdots ,0_{(i-1)},0_{1\hdots (i-1)\overline{(i+1)\hdots(L+1)}},0_{(i+1)},\hdots ,0_{(L+1)})\,,
\end{equation}
where we introduce the \textit{dual representation} of an MLT$(L)$ topology, and the arrow is used to indicate that the expression in the r.h.s. is the result of applying CRT to the original amplitude\footnote{As we already mentioned in the previous discussion, all the formulae presented here are valid for integrands with non trivial numerators. So, we can directly promote $G_F \rightarrow {\cal A}_F$ and $G_D \rightarrow {\cal A}_D$, for the original Feynman and dual integrands, respectively, of scattering amplitudes.}. As pictorially depicted in Fig.~\ref{fig:MLTdual}, this formally proves the validity of the MLT$(L)$ formulae presented in Ref.~\cite{Verdugo:2020kzh}.

In the specific case of MLT$(L)$ topologies with single powers and one propagator per loop set, we formally proof the formulae presented in Ref.~\cite{Verdugo:2020kzh}. After summing over all the dual $L+1$ contributions, and applying the results given in Appendix~\ref{app:simpdualexp}, Eq.~(\ref{eqn:5.1}) collapse to the extremely compact and causal expression
\begin{equation}
    G_F(1,\hdots,L+1)\rightarrow-\frac{1}{\prod\limits_{k=1}^{L+1}\left(2q_{k,0}^{(+)}\right)}\left(\frac{1}{\sum\limits_{k=1}^{L+1}q_{k,0}^{(+)}-k_{L+1}}+\frac{1}{\sum\limits_{k=1}^{L+1}q_{k,0}^{(+)}+k_{L+1}}\right).
\label{eqn:5.5}\end{equation}This is a multi-loop generalization of Eq.~(\ref{eq:ResultadoEjemplito}).

It is worth mentioning that Eq.~(\ref{eqn:5.5}) is a consequence of the algebraic properties of the nested residues, and the same strategy can be applied in order to show the explicit causal representations for more complex topologies as exhibited in Ref.~\cite{Aguilera-Verdugo:2020kzc}.

\subsection{NMLT$(L)$ and N$^2$MLT$(L)$}
\label{ssec:NMLT+NNMLT}
The MLT configuration is sufficient to describe any two-loop scattering amplitude, but new mathematical structures appear at higher orders. Starting at three loops, we also need to consider the NMLT and N$^2$MLT topologies.  In fact, the NMLT topology is described as a subtopology of N$^2$MLT, which is the master topology at three loops. This constitutes a clear and powerful classification scheme towards an efficient computation of higher-order amplitudes, since these new topologies involve new loop momenta linear combinations. The NMLT$(L)$ and N$^2$MLT$(L)$ are respectively characterised as follows;
\begin{align}
\mathcal{A}_{\mathrm{NMLT}}^{(L)}(1,\hdots,L+2)&\equiv\int\limits_{\ell_1,\hdots,\ell_L}\mathcal{N}(\{\ell_i\}_L,\{p_j\}_N) \times G_F(1,\hdots,L+2)\,,\\
\mathcal{A}_{\mathrm{N}^2\mathrm{MLT}}^{(L)}(1,\hdots,L+3)&\equiv\int\limits_{\ell_1,\hdots,\ell_L}\mathcal{N}(\{\ell_i\}_L,\{p_j\}_N)\times G_F(1,\hdots,L+3)\,,
\label{eqn:n2mltamp}\end{align}
where we include two extra sets, i.e. $\{L+2,L+3\}$, with the aim of describing all the possible momenta configurations.

\begin{figure}[H]
    \centering
    \includegraphics[width=0.32\textwidth]{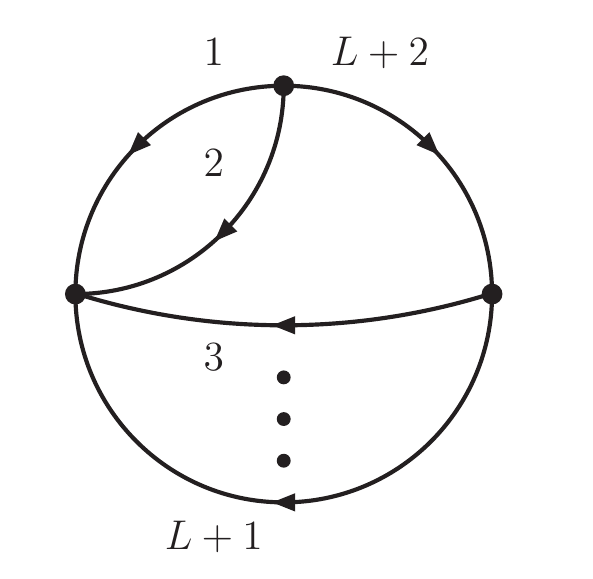}
    \caption{The general form of the NMLT$(L)$ topology.}
    \label{fig:NMLTdiagram}
\end{figure}

\begin{figure}[H]
    \centering
    \includegraphics[width=0.65\textwidth]{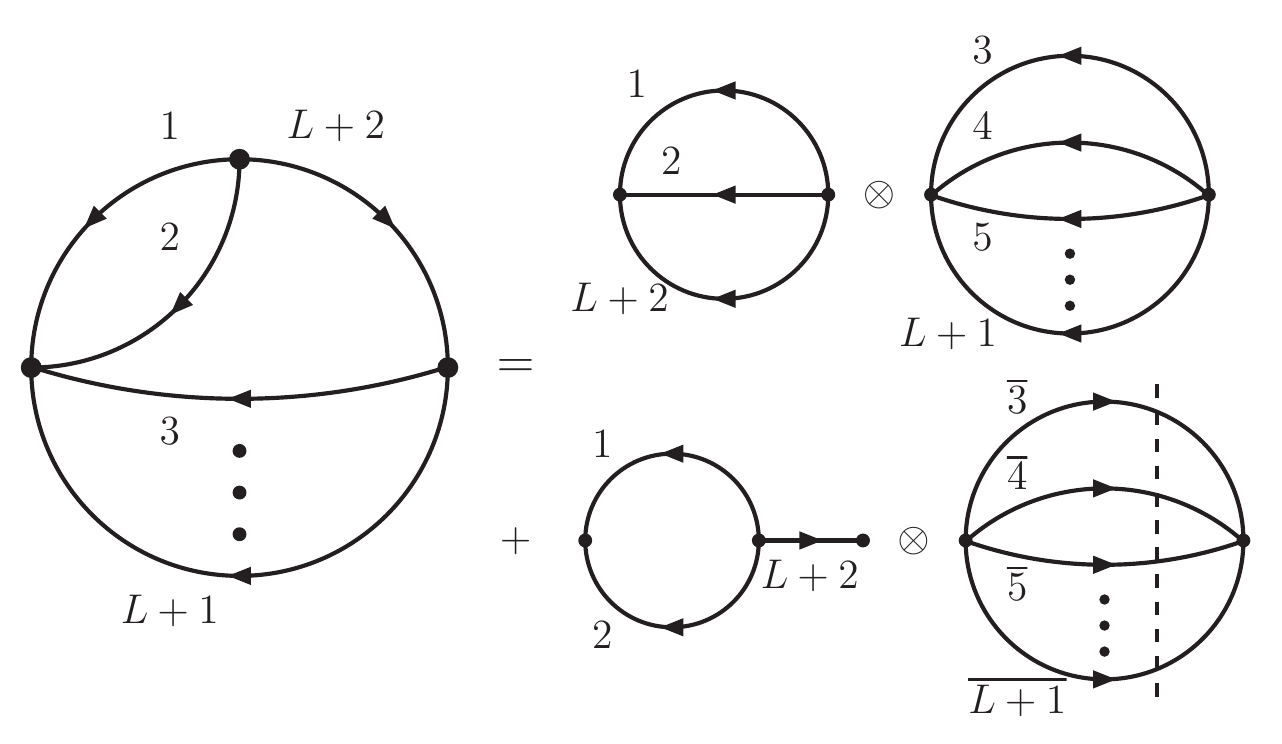}
    \caption{Dual decomposition of NMLT$(L)$ in terms of loop configurations with lower topological complexity.}
    \label{fig:NMLTdual}
\end{figure}

We begin by describing the NMLT$(L)$ vacuum diagram, which is pictorially shown in Fig.~\ref{fig:NMLTdiagram}. 
To simplify the notation, the additional set $L+2=\overline{12}$ only 
contains the linear combination of the two loop momenta $\ell_1$ and $\ell_2$. Regarding this case, Eq.~(\ref{eqn:38}) can be rewritten in terms of Feynman propagators and its dual expansion as 
\begin{equation}\begin{split}
    G_F(1,\hdots,L+2)&\rightarrow G_D(1,2,L+2)\otimes G_{3\hdots(L+1)}(L+2^*)\\
    &+G_D(1,2)\otimes G_{F}(L+2)\otimes G_{3\hdots(L+1)}(0_{(3\hdots(L+1))}^*). \,
\end{split}
\label{eqn:5.2}
\end{equation}
This expression can be understood in terms of loop configurations of lower topological complexity as it is shown in Fig.~\ref{fig:NMLTdual}. As it was anticipated in Sec.~\ref{sec:notation}, the convolution symbol is not a pure factorization. We would like to emphasize that it implies the use of the on-shell conditions to express all the off-shell variables.

We observe that the NMLT$(L)$ displays two contributions in terms of MLT configurations. Explicitly, the first contribution is a convolution of an MLT$(L-2)$ with an MLT$(2)$ diagram, and the second term is a convolution of an MLT$(L-2)$ diagram, all of its propagators are on shell and are reversed, with an MLT$(1)$ one.

\begin{figure}[H]
    \centering
    \includegraphics[width=0.32\textwidth]{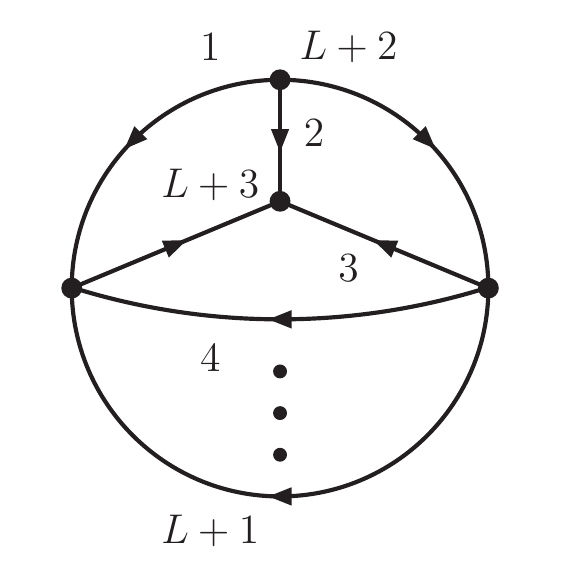}
    \caption{N$^2$MLT$(L)$ diagram with $L$ loops.}
    \label{fig:NNMLTdiagram}
\end{figure}

Finally, let us now draw our attention to the $\mathrm{N}^2\mathrm{MLT}(L)$ configurations. It is worth appreciating that this mathematical object contains the highest topological complexity at three-loop level. In order to describe it, we need to add the set $L+3 = \overline{23}$, which is the set that can only contain combinations of the loop momenta $\ell_2$ and $\ell_3$. The generated vacuum topology, or Mercedes-Benz like-diagram, is depicted in Fig.~\ref{fig:NNMLTdiagram}.

\begin{figure}[H]
    \centering
    \includegraphics[width=0.65\textwidth]{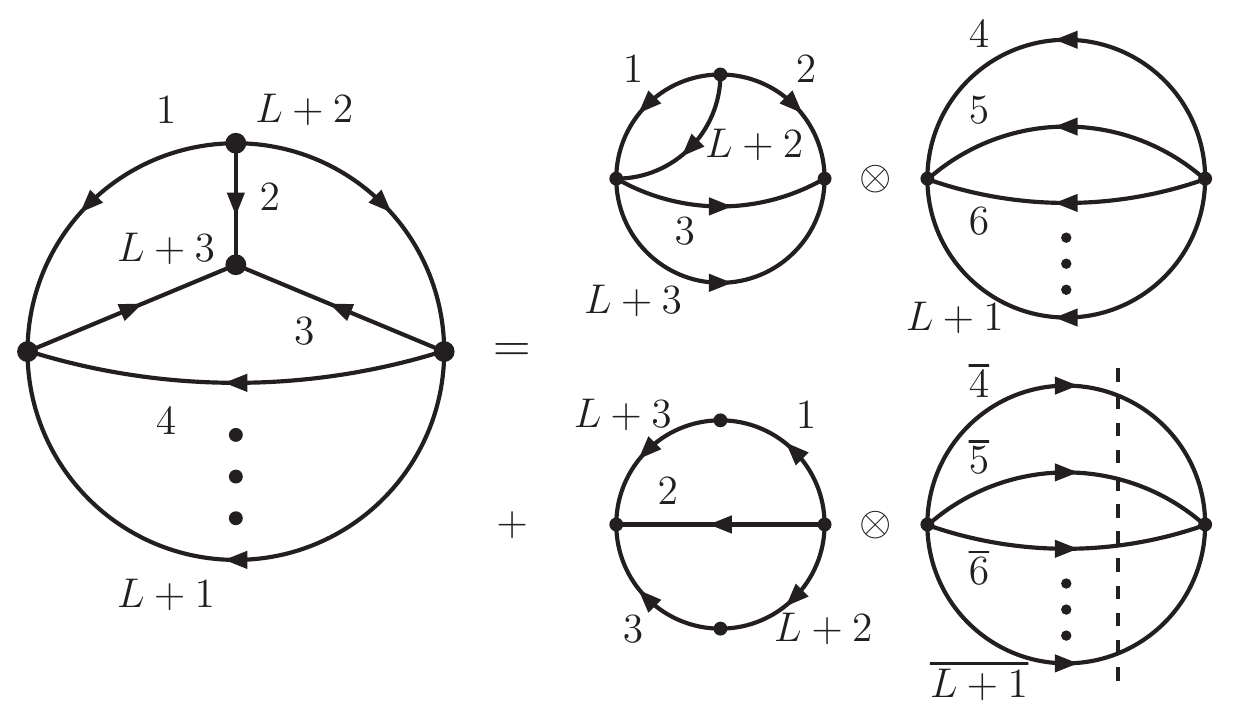}
    \caption{Dual expansion of a N$^2$MLT$(L)$ diagram.}
    \label{fig:NNMLTdual}
\end{figure}

In order to find the dual expression of Eq.~(\ref{eqn:n2mltamp}), we proceed to apply the CRT consecutively and, after computing the first residue and taking into account Eq.~\eqref{eqn:3.30}, we obtain, \begin{equation}
    \begin{split}
    G_F(1,\hdots,L+3)\rightarrow&G_D(1,2,3,L+2,L+3)\otimes G_{D}(0_{(4\hdots(L+1)))}^*)\\
    +&G_D(\langle1,L+3\rangle,2,\langle L+2,3\rangle)\otimes G_{F}(123^*) .
    \end{split}
    \label{eq:decon2mlt}
\end{equation} 
This decomposition is graphically described in Fig.~\ref{fig:NNMLTdual}, where the convolution symbols have the same interpretation as before. The brackets notation used for $G_D(\langle 1,L+3\rangle,2,\langle L+2,3\rangle)$ corresponds to the insertion of external momenta in specific internal lines. For instance, let us take a look at the second line of Fig.~\ref{fig:NMLTdual}, where the dot between $1$ and $L+3$ represents the insertion of an \emph{external} particle with momenta $(L+3)\overline{1}$. It is also important to remark that we obtain a dual expansion with two contributions where one of the terms is a convolution of an MLT$(L-3)$ configuration with an NMLT$(3)$, and the other term consists of one MLT$(L-3)$, all of its internal momenta set on shell and reversed, and an MLT$(2)$ configuration with two propagators in two internal lines. 

The decompositions of NMLT$(L)$ and N$^2$MLT$(L)$ topologies shows explicit recursion relations involving configurations with lower topological complexity. Therefore, from the above studies and by an iterated application of the decomposition presented in Figs.~\ref{fig:NMLTdual} and~\ref{fig:NNMLTdual}, we can notice that any N$^{k-1}$MLT (for $k\leq3$) can be cast in terms of MLT configurations. 
Besides, it is interesting to point out that recent studies that include master topologies at four-loop also present the same behaviour~\cite{Ramirez-Uribe:2020hes}. Therefore, an extensive study of MLT configurations and their convolutions, as carried out in the present paper, is sufficient to understand the behaviour of any $L$-loop amplitude with any number of external legs.

\subsection{Higher topological complexity and causality}
\label{ssec:highertopo}
The ideas presented in this paper can be generalized to scattering amplitudes with an arbitrary topological complexity. This is due to the fact that the algorithm for the computation of the nested residue does not depend on the number of loops nor the topological classification of the diagram. Moreover, the algorithmic procedure is the same whether or not external particles are present.

\begin{figure}[H]
    \centering
    \includegraphics[width=0.75\textwidth]{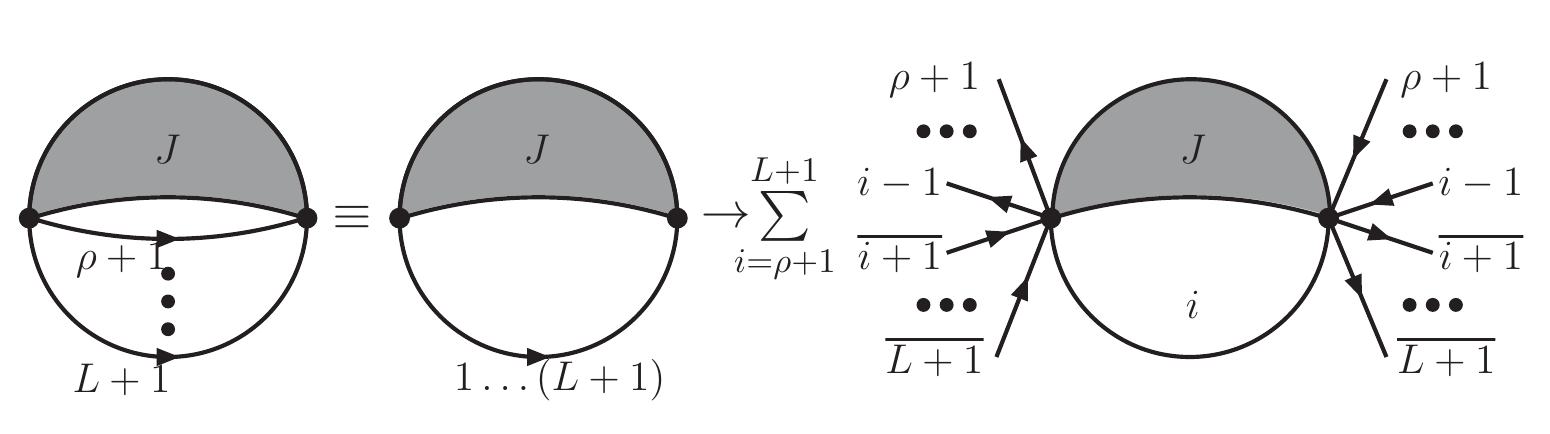}
    \caption{Graphical factorization of a multi-loop topology with a MLT insertion. The gray blob represents the subtopology with specific topological complexity. The diagram on the r.h.s. of the arrow represents the opening of the MLT subtopology.}
    \label{fig:8}
\end{figure}

In order to deal with an amplitude with arbitrary topological complexity, it is useful to express Eq.~(\ref{eqn:simp1}) as it is shown in Fig.~\ref{fig:8}, where $\alpha=1\hdots\rho^*$ plays the role of the internal line $L+1$ in the minimal Feynman diagram with the same topological complexity. In other words, the line $\alpha$ is equivalent to unifying all the MLT-like insertions of the original diagram In this figure, the topological complexity of the diagram has been isolated inside the blob $J$, and the remaining MLT-like part of the diagram is simplified with the results of this work. This is important because, after the computation of the iterated residue (equivalently, after a partial opening the diagram), the presence of external particles attached to the vertices isolating the topological complexity can be thought as merged into a new internal line, whose momenta flow is determined by momentum conservation. We would like to highlight that Eq.~(\ref{eqn:simp1}) nor Fig.~\ref{fig:8} are final results, since still remains the nested residues with respect to momenta 1, 2, ..., $\rho$ and $\alpha$ have to be computed. 

Also, for interactions of external particles with the internal lines of the subdiagram $J$, the analytical structure remains untouched: all the information regarding the external lines is codified inside the factorized contribution associated with $J$. Thus, we can only think about a different $J'$, clearly with additional poles, but the factorization formula remains the same.

Some works have been developed for the study of the N$^3$MLT($L$) and N$^4$MLT($L$)~\cite{Ramirez-Uribe:2020hes}. There, it is seen that the computation of the nested residue for topological complexities 4 and 5 yields to representations analogous to Eqs. (\ref{eqn:5.2}) and (\ref{eq:decon2mlt}).

\begin{figure}[H]
    \centering
    \includegraphics[width=0.97\textwidth]{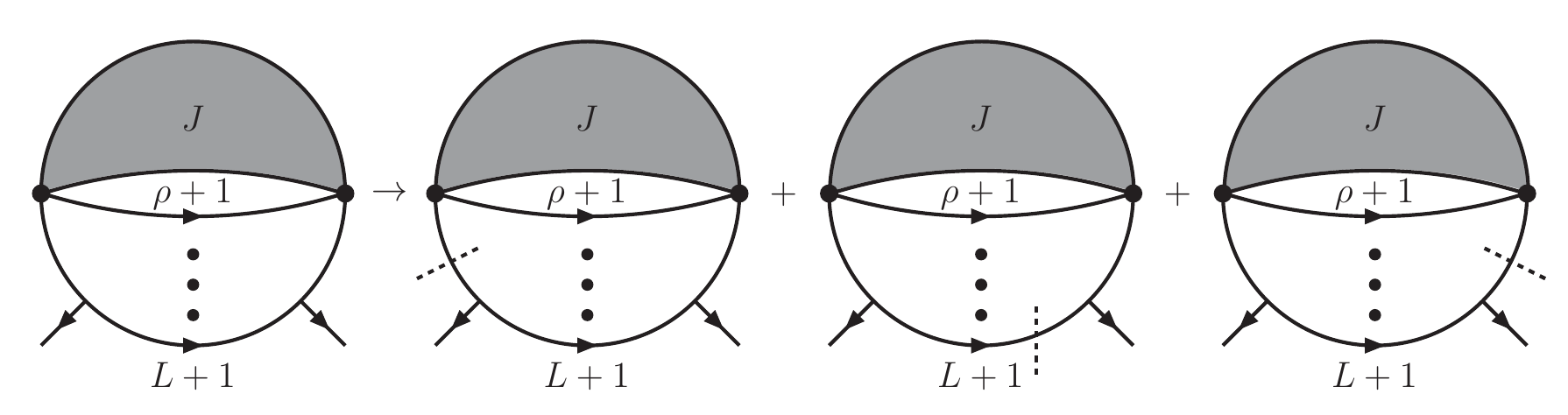}
    \caption{Arbitrary topological complexity with external particles.}
    \label{fig:9}
\end{figure}

If external particles are attached to an internal line of the MLT subtopology, the sum over the nested residues of all the propagators that belong to the corresponding set is required. All the poles within each set have imaginary parts of the same sign. In this situation, the generalization of the results derived from Eq.~(\ref{eqn:snmlt}) becomes straightforward: the nested residue takes the form of a sum of as many copies as propagators are in the same internal line, where the poles are shifted one to another by a real number. This configuration is depicted in Fig.~\ref{fig:9}.

The application of the ideas presented in this paper will be useful when a realistic scattering process is considered; where it will become mandatory to study the consequences of having a polynomial in the energy components of the loop momenta $q_{i,0}$ as numerator. For the purposes of this work, it is not necessary to make an explicit example to claim that when the numerator is not identically 1, the results presented along this document are still valid if the numerator is a meromorphic function in every energy variable. And, since this is the case for a Feynman integral (the integrand is always a rational function of the energy of the loop momenta), then these results stand for any QFT. In addition, this approach can be used to obtain the causal structure of an arbitrary topological class of diagrams, since the nested residue leads, in a natural way, to sums of on-shell energy of the internal particles, $q_{k,0}^{(+)}$ (similar to the expressions of Appendix~\ref{app:simpdualexp}), avoiding the non-physical threshold singularities. These causal structures makes it easier to localize the physical thresholds, as they will play an explicit role within the causal denominators.

Finally, we would like to make a brief comment about causality and the location of physical thresholds. As discussed in Ref.~\cite{Aguilera-Verdugo:2020kzc}, when adding up all the dual contributions, the resulting expression is written in terms of \textit{causal denominators}, $\{\lambda_i^{\pm}\}$. These variables represent sums of on-shell energies and combinations of the energy of the external particles. The number of causal denominators depends on the topological complexity and on the number of external particles. However, their functional form and explicit dependence on the number of loops can be inferred from the causal denominators present in the associated vacuum diagram. Thus, the proofs provided in this article allow to ensure the validity of the all-loop order formulae presented in Ref.~\cite{Aguilera-Verdugo:2020kzc}.

\section{Conclusions}
\label{sec:conclusions}
The computation of scattering amplitudes at higher-orders in the perturbative expansion is a very challenging task, specially for multi-leg processes. Even though several highly innovative and groundbreaking techniques were developed in recent years, automation of multi-loop scattering amplitudes still remains a frontier problem. In this respect, the Loop-Tree Duality offers alternative representations of generic scattering amplitudes at integrand level, which have many potential advantages over the customary Feynman representation.

In this paper, we deepened into the mathematical aspects of the multi-loop construction presented in Ref.~\cite{Verdugo:2020kzh}. Firstly, we provided a rigorous definition of the multi-iterated residue computation using a generic test function and exploring the consequences of the prescription introduced. We found that it is not necessary to keep the residues of all the poles whilst performing the iteration, since some of them cancel. These \emph{displaced poles} are associated to non-physical contributions, which cannot be interpreted in terms of cut diagrams. This allows to redefine integrand representations in terms of the so-called \emph{nested residues} which are only those related to physical contributions.

Inspired by the mathematical properties of the nested residues, we defined a closed notation to achieve an efficient symbolic handling of intermediate expressions. Following this approach, carefully explained in Sec.~\ref{sec:notation}, the residue computation is performed by lowering indices, which contain the relevant information about the pole location. In addition, the nested application of the Cauchy residue theorem leads to recursive structures, whose behaviour is accurately captured by this short-hand notation. 

Once the formalism presented here was provided with a physical meaning, in Sec.~\ref{sec:connectionQFT}, we managed to apply it to specific benchmark amplitudes. In particular, we introduced a formal definition for the topological complexity of families of loop diagrams. We used these concepts and the operational methodology presented in Sec.~\ref{sec:notation}, to inquire into the recursive structure of MLT$(L)$, NMLT$(L)$ and N$^2$MLT$(L)$ topologies. In this way, we provided the ingredients required for an inductive proof of the all-loop order formulas presented in Refs.~\cite{Verdugo:2020kzh}. Moreover, we showed that the recursive nature of the computations leads to an explanation to the causal behaviour of the compact formulae found in previous papers~\cite{Aguilera-Verdugo:2020kzc}. As detailed in Sec.~\ref{ssec:highertopo}, the ideas that we developed can be straightforwardly applied to any multi-loop multi-leg amplitude, independently of their topological complexity~\cite{Ramirez-Uribe:2020hes}.

In summary, we exhaustively focused on the analysis of the mathematical structures behind scattering amplitudes by applying the LTD framework. From the formal properties that we found, we obtained valuable information for proving explicit all-order formulas, and also to efficiently perform the symbolic handling of the expressions. This knowledge allows to reach higher-perturbative orders, thus opening an interesting path for more precise theoretical predictions.

\section*{Acknowledgements}
We gratefully acknowledge Selomit Ram\'irez-Uribe for interesting discussions and graphical support. This work is supported by the Spanish Government (Agencia Estatal de Investigaci\'on) 
and ERDF funds from European Commission (Grant No. FPA2017-84445-P), 
Generalitat Valenciana (Grant No. PROMETEO/2017/053) and the COST Action CA16201 PARTICLEFACE. 
R.J.H.-P. acknowledges support from Departament de F\'isica Te\`orica, Universitat de Val\`encia, CONACyT through
the Project No. A1-S-33202 (Ciencia B\'asica) and Sistema Nacional de Investigadores;
W.J.T.  from Juan de la Cierva program (FJCI-2017-32128), 
and J.J.A.V. from Generalitat Valenciana (GRISOLIAP/2018/101).

\appendix
\section{Cancellation of residues from displaced poles}
\label{app:Cancelacion}
The cancellation of the residues from displaced poles, defined in Sec.\ref{sec:multiLTD}, is guaranteed by the following:

\textit{Lemma}: Let $P(x_i,x_j)$ be a meromorphic function in both variables $x_i$ and $x_j$ whose poles are not located on $\{x_i,y_i+k_i\}$, $\{x_i,y_k-x_j+k_{ij}\}$ nor $\{x_j,y_k-y_i+k_{ij}-k_i\}$, with $k_i,k_{ij},y_i,y_k\in\complex$ where $y_i,y_k\in\{\mathrm{Im}(z)<0\}$, and let \begin{equation}
    F(x_i,x_j)=\frac{P(x_i,x_j)}{((x_i-k_i)^2-y_i^2)^{\gamma_i}((x_i+x_j-k_{ij})^2-y_k^2)^{\gamma_k}}.
\end{equation} Then, the iterated residue in each of the explicit poles satisfies \begin{equation}
    \begin{split}
        &\mathrm{Res}(\mathrm{Res}(F(x_i,x_j),\{x_i,y_i+k_i\}),\{x_j,y_k-y_i+k_{ij}-k_i\})=\\
        -&\mathrm{Res}(\mathrm{Res}(F(x_i,x_j),\{x_i,y_k-x_j+k_{ij}\}),\{x_j,y_k-y_i+k_{ij}-k_i\}).
    \end{split}
\end{equation}

\textit{Proof}: If the shifts $x_i'=x_i-k_i$ and $x_j'=x_j-k_{ij}+k_i$ are performed, the function $F$ can be rewritten in the form \begin{equation}\label{eqn:Fsimple}
    F(x_i',x_j')=\frac{P(x_i',x_j')}{(x_i'^2-y_i^2)^{\gamma_i}((x_i'+x_j')^2-y_k^2)^{\gamma_k}}.
\end{equation}
Without loss of generality, this is also equivalent to consider $k_i=k_{ij}=0$.

The function in Eq.~(\ref{eqn:Fsimple}) has two explicit poles of order $\gamma_i$ and $\gamma_k$ within the half plane $\mathrm{Im}(z)<0$. Thus, the function $F$ has an expansion of the form \begin{equation}\label{eqn:53}
    F(x_i',x_j')=\sum\limits_{r_i=-\gamma_i}^{\infty}\sum\limits_{r_k=-\gamma_k}^{\infty}a_{r_i,r_k}(x_i'-y_i)^{\gamma_i}(x_i'+x_j'-y_k)^{r_k}.
\end{equation}
If the last factor of the right hand side of Eq.~(\ref{eqn:53}) is rewritten in the form $x_i'+x_j'-y_k=(x_i'-y_i)+(x_j'-y_k+y_i)$, and if the sum over $r_k$ is split into negative and non-negative values, it is obtained \begin{equation}\label{eqn:54}
    \begin{split}
        F&(x_i',x_j')=\sum\limits_{r_i=-\gamma_i}^{\infty}\sum\limits_{r_k=-\gamma_k}^{\infty}a_{r_i,r_k}(x_i'-y_i)^{\gamma_i}(x_i'+x_j'-y_k)^{r_k}\\
        &=\sum\limits_{r_i=-\gamma_i}^{\infty}\sum\limits_{r_k=1}^{\gamma_k}\sum\limits_{s=0}^{\infty}(-1)^{r_k+s}a_{r_i,-r_k}\frac{(x_i'-y_i)^{r_i+s}}{(r_k-1)!}(x_j'-y_k+y_i)^{-r_k-s}\prod\limits_{t=1}^{r_k-1}(s+t)\\
        &+\sum\limits_{r_i=-\gamma_i}^{\infty}\sum\limits_{r_k=0}^{\infty}\sum\limits_{s=0}^{r_k}a_{r_i,r_k}\binom{r_k}{s}(x_i'-y_i)^{r_i+s}(x_j'-y_k+y_i)^{r_k-s}.
    \end{split}
\end{equation}
To compute the first residue of the function in Eq.~(\ref{eqn:53}), for $\{x_i',y_i\}$, it is enough to take the coefficient of the term with the factor $(x_i'-y_i)^{-1}$ in the expansion of Eq.~(\ref{eqn:54}). Afterwards, to obtain the second residue, for $\{x_j',y_k-y_i\}$, we select the coefficient of the term with the factor $(x_j'-y_k+y_i)^{-1}$. For the second sum, the second condition is never satisfied because $0\leq s\leq r_k$ and such a factor demands the condition $s=r_k+1$. For the first sum, the second condition is obtained for $r_k+s=1$, and as $1\leq r_k\leq\gamma_k$ and $0\leq s<\infty$, there is just one term satisfying this condition, with $s=0$ and $r_k=1$. As $s=0$, the first condition is satisfied for $r_i=-1$. Hence \begin{equation}
    \mathrm{Res}(\mathrm{Res}(F(x_i',x_j'),\{x_i',y_i\}),\{x_j',y_k-y_i\})=-a_{-1,-1}.
\end{equation}
If in the function in Eq.~(\ref{eqn:53}), we rewrite the second factor as $x_i'-y_i=(x_i'+x_j'-y_k)-(x_j'-y_k+y_i)$, and if the sum over $r_i$ is split into negative and non-negative values, we obtain \begin{equation}
    \begin{split}\label{eqn:58}
        F(x_i',x_j')&=\sum\limits_{r_i=1}^{\gamma_i}\sum\limits_{r_k=-\gamma_k}^{\infty}\sum\limits_{s=0}^{\infty}a_{-r_i,r_k}\frac{(x_i'+x_j'-y_k)^{r_k+s}}{(r_i-1)!}(x_j'-y_k+y_i)^{-r_i-s}\prod\limits_{t=1}^{k_1-1}(s+t)\\
        &+\sum\limits_{r_i=0}^{\infty}\sum\limits_{r_k=-\gamma_k}^{\infty}\sum\limits_{s=0}^{r_i}a_{r_i,r_k}\binom{r_i}{s}(x_i'+x_j'-y_k)^{r_k+s}(-x_j'+y_k-y_i)^{r_i-s}.
    \end{split}
\end{equation}
Again, the iterated residue of this expression is the coefficient of the terms proportional to $(x_i'+x_j'-y_k)^{-1}$ and $(x_j'-y_k+y_i)^{-1}$. For the first sum, this conditions are satisfied for $r_k+s=-1$ and $r_i+s=1$. However, as $1\leq r_i\leq \gamma_i$ and $0\leq s$, the last condition is fulfilled only for $r_i=1$ and $s=0$. Thus, the first condition is expressed as $r_k=-1$. For the second sum, the first condition holds, but the second condition shall be expressed as $r_i-s=-1$ so that $s=r_i+1$. However, it is given that $0\leq s\leq r_i$ and then this sum does not contribute to the residue. Thus,\begin{equation}
    \mathrm{Res}(\mathrm{Res}(F(x_i',x_j'),\{x_i',y_k-x_j'\}),\{x_j',y_k-y_i\})=a_{-1,-1}.
\end{equation}
It is then concluded that \begin{equation}
    \begin{split}
        &\mathrm{Res}(\mathrm{Res}(F(x_i',x_j'),\{x_i',y_i\}),\{x_j',y_k-y_i\})=\\
        -&\mathrm{Res}(\mathrm{Res}(F(x_i',x_j'),\{x_i',y_k-x_j'\}),\{x_j',y_k-y_i\}).\ \ \ 
    \end{split}
\end{equation}

If we then restore the original variables that are shifted by $k_i$ and $k_{ij}$ with respect to $x_i'$ and $x_j'$, we arrive to the expression we wanted to demonstrate \begin{equation}\label{eqn:64}
    \begin{split}
        &\mathrm{Res}(\mathrm{Res}(F(x_i,x_j),\{x_i,y_i+k_i\}),\{x_j,y_k-y_i+k_{ij}-k_i\})=\\
        -&\mathrm{Res}(\mathrm{Res}(F(x_i,x_j),\{x_i,y_k-x_j+k_{ij}\}),\{x_j,y_k-y_i+k_{ij}-k_i\}).
    \end{split}
\end{equation}

\section{Proof by induction of the multi-loop MLT($L$) representations}
\label{app:IterativeFormula}
This Appendix presents a formal proof of the dual representation of MLT($L$) in terms of nested residues (Eq.~(\ref{eqn:37})). The proof is given by induction on the number of computed residues through the iterated residues algorithm.

Here, we start by analysing the dual representation of a scalar MLT$(L)$ diagram with one propagator for each set. The original integrand in the Feynman representation is given by \begin{equation}
    \mathcal{I}^{(L)}_{\mathrm{MLT}}=G_F(1,2,\hdots ,L+1)=G_F(1,2,\hdots ,L,1\hdots L).
\end{equation}
After the computation of the first residue with respect to the variable $x_{L}$, we get \begin{equation}\begin{split}
    \mathrm{Res}(\mathcal{I}_{\mathrm{MLT}}^{(L)},\{q_{L,0},\mathrm{Im}(q_{L,0})<0\})&=G_D(1,2,\hdots,L-1,0_{(L)},1\hdots(L-1)_{L})\\
    &+G_D(1,\hdots,n-2,1\hdots(L-1)_{\overline{L+1}},0_{(L+1)}).
\end{split}\end{equation}

In order to prove the cancellation of the contributions of the displaced poles in each iteration of the iterated residues by mathematical induction, we assume that the function obtained after computing the first $i$ iterated residues (for the last $i$ variables) is given by \begin{equation}
    \begin{split}
        &G_F(1,\hdots,L+1)\to G_F(1,\hdots,L-i)\\
        &\times\sum\limits_{j=L-i+1}^{L+1}G_D(0_{(L-i+1)},\hdots,0_{(j-1)},1\hdots(L-i)_{(L-i+1)\hdots(j-1)\overline{(j+1)\hdots(L+1)}},0_{(j+1)},\hdots,0_{(L+1)}),
    \end{split}\label{eqn:hipofind}
\end{equation} where it has been factorized out the Feynman propagator $G_F(1,\ldots,L-i)$ as it depends on independent primitive variables. Then, the set of poles of the function given in Eq.~(\ref{eqn:hipofind}) with respect to the variable $q_{L-i,0}$ is given by \begin{equation}\label{eqn:polesofb3}\begin{split}
    &\mathrm{Poles}[G_F(1,\hdots,L+1),q_{L+1,0},\hdots,q_{L-i-1,0};q_{L-i,0}]\\
    &=\{\pm q_{L-i,0}^{(+)}\}\bigcup\left(\bigcup\limits_{j=L-i+1}^{L}\left\{-\sum\limits_{j=1}^{L-i-1}q_{j,0}\pm q_{j,0}^{(+)}-\sum\limits_{k=L-i+1}^{j-1}q_{k,0}^{(+)}+\sum\limits_{k=j+1}^{L+1}q_{k,0}^{(+)}\right\}\right)\\
    &\bigcup\left\{k_{L+1,0}-\sum\limits_{n=1}^{L-i-1}q_{n,0}-\sum\limits_{n=L-i+1}^{L}q_{n,0}^{(+)}\pm q_{L+1,0}^{(+)}\right\}.
\end{split}\end{equation}

Although the first component in Eq.~(\ref{eqn:polesofb3}) has a single negative-imaginary-part pole, namely $q_{L-i,0}^{(+)}$, the second component contains one negative-imaginary-part pole and the third component has one positive-imaginary-part pole, because, \begin{equation}\label{eqn:definitepoles}
    \begin{split}
        \mathrm{Im}\left(\sum\limits_{k=L-i+1}^{L+1}q_{k,0}^{(+)}\right)&<0,
    \end{split}
\end{equation} while all other poles are displaced poles, we should select only the residues of the non-displaced poles with negative imaginary part.

Following with the next nested residue,  we get,\begin{equation}\label{eqn:71}
    \begin{split}
        &G_F(1,\hdots,L+1)\to G_F(1,\hdots,L-i-1)\\
        &\times\Bigg(\sum\limits_{j=L-i}^{L+1}G_D(0_{(L-i)},\hdots,0_{(j-1)},1\hdots(L-i-1)_{(L-i)\hdots(j-1)\overline{(j+1)\hdots(L+1)}},0_{(j+1)},\hdots,0_{(L+1)})\\
        &+G_D(1\hdots(L-i-1)_{(L-i)\hdots(L+1)},0_{(L-i)}\hdots,0_{(L+1)})\Bigg)\\
        &=G_F(1,\hdots,L-i-1)\\
        &\times\sum\limits_{j=L-i}^{L+1}G_D(0_{(L-i)},\hdots,0_{(j-1)},1\hdots(L-i-1)_{(L-i)\hdots(j-1)\overline{(j+1)\hdots(L+1)}},0_{(j+1)},\hdots,0_{(L+1)}).
    \end{split}
\end{equation}
Hence, this proves by induction that the computation of the first $i$ iterated residues, results into the expression in Eq.~(\ref{eqn:hipofind}).

In particular, after computing all the residues, it is obtained \begin{equation}
    \begin{split}
        G_F(1,\hdots,L+1)&\rightarrow\sum\limits_{j=1}^{L+1}G_D(0_{(1)},\hdots,0_{(j-1)},0_{1\hdots(j-1)\overline{(j+1)\hdots(L+1)}},0_{(j+1)},\hdots,0_{(L+1)}).\ \ \ 
    \end{split}
\end{equation}

It is worth to say that the proof of Eq.~(\ref{eqn:hipofind}) is general enough to cover the case involving an arbitrary topological complexity. This is because we can isolate the higher-topology structure inside the factor $G_F(1,\hdots,L-i-1)$, and proceed as described. Thus, Eq.~(\ref{eqn:hipofind}) can be applied to Feynman diagrams with higher topological complexity and any number of loops.

\section{Causal rearrangement of nested residues}
\label{app:simpdualexp}
Let $\{1,\hdots,\rho\}$ be the family of sets with loop momenta appearing in three or more sets, and let $\rho+1\leq i\leq L$. After the computation of the $i$-th iterated residue of the integrand of a general Feynman integral, \begin{equation}
    \mathcal{A}^{(L)}_{\mathrm{N}^{k-1}\mathrm{MLT}}=\int\limits_{\ell_1,\hdots,\ell_{L}}N(\{\ell_{i}\}_L,\{p_j\}_N)\times G_F(1,\hdots,L+k),
\end{equation}we obtain\begin{equation}
    \begin{split}
        &G_F(1,\hdots,L+k)\rightarrow G_F(1,\hdots,L-i,L+2,\hdots,L+k)\\
        &\times\sum\limits_{j=L-i+1}^{L+1}G_D(0_{(L-i+1)},\hdots,0_{(j-1)},1\hdots(L-i)_{(L-i+1)\hdots(j-1)\overline{(j+1)\hdots(L+1)}},0_{(j+1)},\hdots,0_{(L+1)}),
    \end{split}
\end{equation}
which is a generalization of Eq.~(\ref{eqn:hipofind}). If this expression is written explicitly in terms of dual propagators, then for the simplest case with $\gamma_j=1$, we have \begin{equation}
    \begin{split}
        &G_D(0_{(L-i+1)},\hdots,0_{(j-1)},1\hdots(L-i)_{(L-i+1)\hdots(j-1)\overline{(j+1)\hdots(L+1)}},0_{(j+1)},\hdots,0_{(L+1)})\\
        =&\frac{1}{\prod\limits_{r=L-i+1}^{L+1}\left(2q_{r,0}^{(+)}\right)}\left(\frac{1}{\sum\limits_{\nu=1}^{L-i}q_{\nu,0}+k_{L+1,0}+\sum\limits_{\nu=L-i+1}^{j-1}q_{\nu,0}^{(+)}-\sum\limits_{\nu=j}^{L+1}q_{\nu,0}^{(+)}}\right.\\
        &\phantom{=\frac{1}{\prod\limits_{k=L-i}^{L+1}\left(2q_{k,0}^{(+)}\right)}\Bigg(}\left.-\frac{1}{\sum\limits_{\nu=1}^{L-i}q_{\nu,0}+k_{L+1,0}+\sum\limits_{\nu=L-i+1}^{j}q_{\nu,0}^{(+)}-\sum\limits_{\nu=j+1}^{L+1}q_{\nu,0}^{(+)}}\right).
    \end{split}
\end{equation}

In this way, by summing over the first $L$ dual terms, we obtain \begin{equation}
    \begin{split}
        \sum\limits_{j=L-i+1}^{L}&G_D(0_{(L-i+1)},\hdots,0_{(j-1)},1\hdots(L-i)_{(L-i+1)\hdots(j-1)\overline{(j+1)\hdots(L+1)}},0_{(j+1)},\hdots,0_{(L+1)})\\
        &=\frac{1}{\prod\limits_{k=L-i+1}^{L+1}\left(2q_{k,0}^{(+)}\right)}\left(\sum\limits_{j=L-i+1}^{L}\frac{1}{\sum\limits_{\nu=1}^{L-i}q_{\nu,0}+k_{L+1,0}+\sum\limits_{\nu=L-i+1}^{j-1}q_{\nu,0}^{(+)}-\sum\limits_{\nu=j}^{L+1}q_{\nu,0}^{(+)}}\right.\\
        &\phantom{=\frac{1}{\prod\limits_{k=L-i}^{L+1}\left(2q_{k,0}^{(+)}\right)}\Bigg(}\left.-\sum\limits_{j=L-i+1}^{L}\frac{1}{\sum\limits_{\nu=1}^{L-i}q_{\nu,0}+k_{L+1,0}+\sum\limits_{\nu=L-i+1}^{j}q_{\nu,0}^{(+)}-\sum\limits_{\nu=j+1}^{L+1}q_{\nu,0}^{(+)}}\right).
    \end{split}
\end{equation}
This last expression is a telescopic series, such that \begin{equation}\label{eqn:C.5}
    \begin{split}
        \sum\limits_{j=L-i+1}^{L}&G_D(0_{(L-i+1)},\hdots,0_{(j-1)},1\hdots(L-i)_{(L-i+1)\hdots(j-1)\overline{(j+1)\hdots(L+1)}},0_{(j+1)},\hdots,0_{(L+1)})\\
        &=\frac{1}{\prod\limits_{r=L-i+1}^{L+1}\left(2q_{r,0}^{(+)}\right)}\left(\frac{1}{\sum\limits_{\nu=1}^{L-i}q_{\nu,0}+k_{L+1,0}-\sum\limits_{\nu=L-i+1}^{L+1}q_{\nu,0}^{(+)}}\right.\\
        &\phantom{=\frac{1}{\prod\limits_{r=L-i+1}^{L+1}\Bigg(}}-\left.\frac{1}{\sum\limits_{\nu=1}^{L-i}q_{\nu,0}+k_{L+1,0}+\sum\limits_{\nu=L-i+1}^{L}q_{\nu,0}^{(+)}-q_{L+1,0}^{(+)}}\right).
    \end{split}
\end{equation} For the last term, it is given that \begin{equation}\label{eqn:C.6}
    \begin{split}
        &G_D(0_{(L-i+1)},\hdots,0_{(L)},1\hdots(L-i)_{(L-i+1)\hdots L})\\
        &=\frac{1}{\prod\limits_{r=L-i+1}^{L+1}\left(2q_{r,0}^{(+)}\right)}\left(\frac{1}{\sum\limits_{\nu=1}^{L-i}q_{\nu,0}+k_{L+1,0}+\sum\limits_{\nu=L-i+1}^{L}q_{\nu,0}^{(+)}-q_{L+1,0}^{(+)}}\right.\\
        &\phantom{=\frac{1}{\prod\limits_{r=L-i+1}^{L+1}\left(2q_{r,0}^{(+)}\right)}}-\left.\frac{1}{\sum\limits_{\nu=1}^{L-i}q_{\nu,0}+k_{L+1,0}+\sum\limits_{\nu=L-i+1}^{L+1}q_{\nu,0}^{(+)}}\right)
    \end{split}
\end{equation}

Adding up Eqs.~(\ref{eqn:C.5}) and~(\ref{eqn:C.6}) it is obtained \begin{equation}
    \begin{split}
        \sum\limits_{j=L-i+1}^{L+1}&G_D(0_{(L-i+1)},\hdots,0_{(j-1)},1\hdots(L-i)_{(L-i+1)\hdots(j-1)\overline{(j+1)\hdots(L+1)}},0_{(j+1)},\hdots,0_{(L+1)})\\
        &=\frac{2\sum\limits_{r=L-i+1}^{L+1}q_{r,0}^{(+)}}{\prod\limits_{L-i+1}^{L+1}\left(q_{r,0}^{(+)}\right)}\frac{1}{\left(\sum\limits_{\nu=1}^{L-i}q_{\nu,0}+k_{L+1,0}\right)^2-\left(\sum\limits_{\nu=L-i+1}^{L+1}q_{\nu,0}^{(+)}\right)}\\
        &=:\frac{2\sum\limits_{r=L-i+1}^{L+1}q_{r,0}^{(+)}}{\prod\limits_{L-i+1}^{L+1}\left(q_{r,0}^{(+)}\right)}G_{(L-i+1)\hdots(L+1)}(1\hdots(L-i)^*),
    \end{split}
\end{equation}where it is defined\begin{equation}
    G_{(L-i+1)\hdots(L+1)}(1\hdots(L-i)^*)=\frac{1}{\left(\sum\limits_{\nu=1}^{L-i}q_{\nu,0}+k_{L+1,0}\right)^2-\left(\sum\limits_{\nu=L-i+1}^{L+1}q_{\nu,0}^{(+)}\right)}.
\end{equation}

In the specific case where $\rho$ does not exist, then $i=L$ and we recover Eq.~(\ref{eqn:5.5}) which corresponds to the causal representation of MLT$(L)$~\cite{Verdugo:2020kzh,Aguilera-Verdugo:2020kzc}.
\\

Finally, from the previous discussion, we can formulate the following:

\textit{Corollary}: For $i=\rho+1$, the first $L-\rho$ nested residues of the function $G_F(1,\hdots,L+k)$ leads to \begin{equation}
    \begin{split}
        G_F&(1,\hdots,L+k)\to G_F(1,\hdots,\rho,L+2,\hdots,L+k)G_{(\rho+1)\hdots(L+1)}(1\hdots\rho^*).
    \end{split}
\end{equation}

\section{Topological reduction with auxiliary propagator}
\label{app:alebra}

Throughout this work it was proved that, for a function $F(1,\hdots,\rho,\hdots,L+1,L+2,\hdots,m)$ with simple poles and vanishing $k$-parameters, if $\{1,\hdots,\rho\}$ is the set of indices of the variables appearing in at least 3 sets, the partial nested residue for the variables $\rho+1,\hdots,L$ gives \begin{equation}\label{eqn:d1}
    F(1\,\hdots,m)\to F(1,\hdots,\rho,L+2,\hdots,m)F_{(\rho+1)\hdots(L+1)}(1\hdots\rho^*),
\end{equation}where the function $F_{(\rho+1)\hdots(L+1)}(1\hdots\rho)$ is a propagator-like function of the form \begin{equation}\label{eqn:auxprop}
    F_{(\rho+1)\hdots(L+1)}(1\hdots\rho^*)=\frac{1}{\left(\sum\limits_{k=1}^{\rho}x_{k}\right)^2-\left(\sum\limits_{k=\rho+1}^{L+1}y_{k}\right)^2}.
\end{equation}

It is important to notice that the function in Eq.~(\ref{eqn:auxprop}) is not a physical propagator. This is the reason we call it \emph{auxiliary propagator}.

In order to formalize the sufficiency of the usage of simple poles and no external particles, let\begin{equation}
    \mathcal{F}_{L,m}:=\left.\left\{F(1,\hdots,m)=\prod\limits_{k=1}^{L}(x_k^2-y_k^2)^{-\gamma_k}\prod\limits_{k=L+1}^{m}(z_k^2-y_k^2)^{-\gamma_k}\right|\gamma_k\in\mathbb{N}\right\}\subseteq\complex^{(\real^L)},
\label{eqn:integrandfamily}\end{equation}be the set of all vacuum polarization Feynman integrands with $L$ loops and $m$ internal particles, and with $y_i\neq y_j$ for $i\neq j$, and let \begin{equation}
    G_{L,m}:=\prod\limits_{k=1}^{L}(x_k^2-y_k^2)^{-1}\prod\limits_{k=L+1}^{m}(z_k^2-y_k^2)^{-1},
\label{eqn:Glmdef}\end{equation}be a Feynman integrand with $L$ primitive variables and $m$ factors with simple poles only. It is evident that $G_{L,m}\in\mathcal{F}_{L,m}$. Let us also define the function \begin{equation}\begin{split}
    \psi:\{G_{L,m}\}\times\mathbb{N}^m&\longrightarrow\mathcal{F}_{L,m}\\
    (G_{L,m},(\gamma_1,\hdots,\gamma_m)&)\mapsto\prod\limits_{k=1}^{L}(x_k^2-y_k^2)^{-\gamma_k}\prod\limits_{k=L+1}^{m}(z_k^2-y_k^2)^{-\gamma_k}.
\end{split}\end{equation} It is worth to notice that this function is biyective, so that the inverse image $\psi^{-1}$ is a function.

After the computation of $k$-th iterated residues with respect to $L,L-1,\hdots,L-k+1$ to every element of $\mathcal{F}_{L,m}$ it is obtained the subset $\mathrm{Res}_k[\mathcal{F}_{L,m}]$ of $\complex^{(\real^{L-k})}$. Finally, let us define the operator \begin{equation}
    \begin{split}
        \Phi_{k}:\mathrm{Res}_k[\{G_{L,m}\}]\times\mathbb{N}^m&\longrightarrow\mathrm{Res}_k[\mathcal{F}_{L,m}]\\
        (f,(\gamma_1,\hdots,\gamma_m)&)\mapsto\left(\prod\limits_{\nu=1}^{m}\frac{1}{(\gamma_\nu-1)!}\frac{\partial^{\gamma_\nu-1}}{\partial\left(q_{\nu,0}^{(+)2}\right)^{\gamma_\nu-1}}\right)f.
    \end{split}
\end{equation}

\begin{figure}[H]
    \centering
    \includegraphics[width=\textwidth]{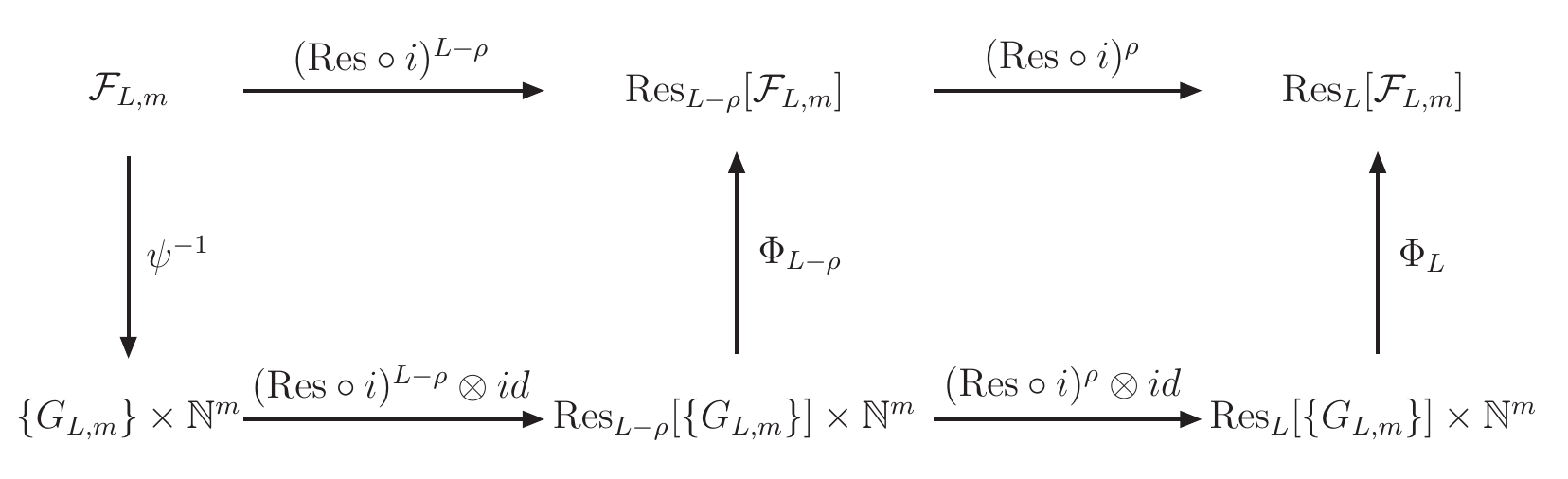}
    \caption{Algebraic diagram that shows the connection among expressions with simple and multiple poles, after computing the nested residues.}
    \label{fig:algebra}
\end{figure}

Using the identity mapping in $\mathbb{N}^m$, $id:\mathbb{N}^m\ni\vec{\gamma}\mapsto\vec{\gamma}\in\mathbb{N}^m$, we show that the algebraic diagram in Fig.~\ref{fig:algebra} commutes.

The proof relies on the fact that, if $F(i)=(x_i^2-y_i^2)^{-1}$, then, \begin{equation}
    \left(F(i)\right)^{\alpha}=\frac{1}{(\alpha-1)!}\frac{\partial^{\alpha_i-1}}{\partial\left(y_{i}^{2}\right)^{\alpha-1}}F(i).
\end{equation}In Ref.~\cite{Aguilera-Verdugo:2020kzc}, this transformation is used to relate the causal structure obtained in Ref.~\cite{Verdugo:2020kzh} for the MLT($L$) with simple Feynman propagators and the expression for a double pole in one of the internal sets. In this work we use it in order to generalize the application to an arbitrary topological complexity, putting on the surface the sufficiency of the simple poles case.

For a general topological complexity diagram, it can be written \begin{equation}\label{eqn:d11}
    F(1^{\gamma_1},\hdots,m^{\gamma_m})=\left(\prod\limits_{k=1}^{m}\frac{1}{(\gamma_k-1)!}\frac{\partial^{\gamma_k-1}}{\partial\left(y_{k}^{2}\right)^{k-1}}\right)F(1,\hdots,m).
\end{equation} Thus, as the derivatives in Eq.~(\ref{eqn:d11}) are not computed with respect to the integration variables $x_{i}$, they commute with the integral. This is, integrating both sides of Eq.~(\ref{eqn:d11}), it is obtained \begin{equation}
    \begin{split}
       \int\limits_{x_1,\hdots,x_L}F(1^{\gamma_1},\hdots,m^{\gamma_m})&=\left(\prod\limits_{k=1}^{m}\frac{1}{(\gamma_k-1)!}\frac{\partial^{\gamma_k-1}}{\partial\left(y_{k}^{2}\right)^{\gamma_k-1}}\right)\int\limits_{x_1,\hdots,x_L}F(1,\hdots,m).
    \end{split}\label{eqn:d9}
\end{equation}
Whence, the computation of the iterated residue of the integrands in both sides of Eq.~(\ref{eqn:d9}) with respect to the same variables in the same order leads to the commutation of the algebraic diagram in Fig.~\ref{fig:algebra}.

\bibliographystyle{JHEP}
\bibliography{refs}

\end{document}